\documentclass[%
preprint,
nofootinbib,
 amsmath,amssymb,
 aps,
pre,
]{revtex4-1}

\usepackage{graphicx}
\usepackage{dcolumn}
\usepackage{bm}

\usepackage[utf8]{inputenc} 
\usepackage[T1]{fontenc}

\usepackage{graphicx} 
\usepackage{epstopdf}
\usepackage[caption=false]{subfig}

\usepackage{float}

\usepackage[inline]{enumitem}
\usepackage{hyperref}
\hypersetup{
         colorlinks   = true,
         allcolors    = blue
          }
\usepackage{booktabs}

\newcommand\etal{\textsl{et al.}} 
\usepackage{cancel}
\usepackage{color}   
\usepackage{framed}
\usepackage{soul}

\usepackage{xcolor}

\AtBeginDocument{%
    \newwrite\bibnotes
    \def\bibnotesext{Notes.bib}
    \immediate\openout\bibnotes=\jobname\bibnotesext
    \immediate\write\bibnotes{@CONTROL{REVTEX41Control}}
    \immediate\write\bibnotes{@CONTROL{%
    apsrev41Control,author="08",editor="1",pages="1",title="0",year="1"}}
     \if@filesw
     \immediate\write\@auxout{\string\citation{apsrev41Control}}%
    \fi
}%
\begin{document}

\title{Large-scale motions from a direct numerical simulation of a turbulent boundary layer}

\author{Ilkay Solak\textsuperscript{1,3}}
\author{Jean-Philippe Laval\textsuperscript{2,}}%
 \email{jean-philippe.laval@univ-lille1.fr}
\affiliation{%
    {\mbox{\textsuperscript{1}Univ. Lille, \textsuperscript{2}CNRS, ONERA, \textsuperscript{3}Centrale Lille, Arts et Metiers ParisTech}} \\
                {Laboratoire de Mécanique des Fluides de Lille - Kampé de Fériet} \\
                {F-59000, Lille, France}}

\begin{abstract}
    A study of large-scale motions from a new direct numerical simulation database of the turbulent boundary layer up to $Re_\theta \sim 2500$ is presented. The statistics of large-scale streamwise structures have been investigated using two-dimensional and three-dimensional extraction procedures. The large-scale structures are abstracted using a robust skeletonization method usually applied to other research domains to simplify complex 3D objects. Different structure parameters such as the length, shape or angle are investigated. The features of the detected structures are compared to their mean counterparts extracted from two-point correlations. Structures as large as 10 boundary layer thickness are observed. The streamwise length of these structures follows a $\text{-}2$ power law distribution, similar to the experimental findings at higher Reynolds numbers. 
\end{abstract}


\maketitle 

\section{Introduction}

Despite numerous recent experiments and numerical simulations, many questions remain unanswered to understand and model the physics of the wall-bounded turbulent flows. The principal reason behind the study of coherent motions is to find analytic relations of the structures corresponding to the significant part of the turbulence statistics. The existence of the coherent structures within the wall-bounded turbulent flows is widely accepted \cite{robinson1991}. Even though different sources show that turbulent boundary layer (TBL) encloses various coherent structures, the puzzle is how these somehow persistent motions can be used to represent boundary layer flows. They can be quickly spotted in most cases (e.g.\, from the images of velocity fluctuations), but a rigorous description of coherent structures is still missing to propose a realistic model of wall-bounded turbulent flows.

The viscous and buffer regions have been studied extensively, and statistical organizations of the coherent structures in this region have been proposed in the literature \cite{adrian2000, lin2006thesis, lin2008}. A recent review of the near-wall region was proposed by Stanislas \cite{stanislas2017}. New investigations on high and very high Reynolds number wall-bounded turbulent flows highlighted the presence of large structures with increasing intensity as the Reynolds number increases. Thus, comprehensive knowledge concerning the dynamics of these large energetic structures might eventually help to model and control high Reynolds number wall-bounded flows. 

These structures usually categorized into large-scale motion (LSM) and very large-scale motion (VLSM) are elongated regions of streamwise velocity fluctuations having a streamwise extent up to 3 boundary layer thickness ($\delta$) for the former and longer than that for the latter\cite{kim1999, guala2006, balakumar2007}. They are located in the logarithmic and lower wake regions of a turbulent boundary layer \cite{hutchins2007b, dennis11b, lee2011}. Special focus has been laid on LSMs and VLSMs as they contribute to a significant amount of turbulent kinetic energy and Reynolds shear stress \cite{ganapathisubramani2003, jimenez04b, wu14, lee2014}. 

Mathis \etal~\cite{mathis2009} have shown that these large-scale structures have an impact on the very near wall statistics starting from the streamwise Reynolds stress associated with the near wall streaks. However, the energy transfer within the full boundary layer thickness is affected by the increasing intensity of these structures. Thus, LSMs and VLSMs have been interpreted to be responsible for the $k_x^{-q}$ scaling range with $q\simeq 1$ of the streamwise velocity spectrum \cite{smits11} and thought to be linked to the attached eddies discussed by Townsend \cite{townsend1956}.

Experimental investigations of the wall-bounded turbulent flows have been conducted mostly based on one point measurements. Optical metrology techniques open some new possibilities for the inspection of the coherent motions in turbulent flows. Even though measurement techniques such as particle image velocimetry (PIV) and Tomo-PIV are promising to access three-dimensional (3D) data, they are not able to reach sufficient resolution on a large size domain to investigate the interaction between the near wall small scales and these large scales. 
Both single point measurements or plane measurements are not able to provide a real 3D shape and arrangement of these very large-scale structures as they rely on Taylor's hypothesis and spanwise regularity. Direct numerical simulation (DNS) allows us to investigate their complex 3D shape and behavior down to the buffer region at high resolution. Meanwhile, numerical simulations of the turbulent boundary layer flow at significant Reynolds numbers are now accessible due to the constant increase of available computational resources. Both spatial and temporal data are essential for the study of coherent structures, and DNS is the only numerical tool to generate high-resolution data in both space and time. 

Few DNS of wall-bounded flows at significant Reynolds numbers has already been performed. Large-scale structures in channel flows have been investigated in detail by Sillero \etal~\cite{sillero2014} using two-point correlations. Their 3D representation of the correlation functions illustrates the potential shape of the streamwise fluctuations structures. Lee \etal~\cite{lee2014} investigated the large-scale motions in channel flows with a detection method which consists of several operations on the velocity fluctuation fields. From temporal analyses, they related the merging of the LSM as a production mechanism for VLSM. These studies and many others reported very large-scale structures up to $15\delta$. LSM up to $20h$ long ($h$ being the channel half-width) have also been observed in channel flows by Del {\'A}lamo \etal~\cite{delalamo2003, delalamo2004}. Hwang \etal~\cite{hwang16} investigated the inner-outer interactions of negative and positive large-scale structures in channel flow at moderate Reynolds number ($Re_\tau=930$) using conditional correlation. In boundary layers, Hutchins and Marusic \cite{hutchins2007b} used hot-wire rake measurements of an atmospheric surface layer and found very long meandering structures up to $20 \delta$  populating the log layer. On the other hand, when viewed from single point statistics, the meandering tendency masks the true length of these structures resulting in shorter length scales. DNS of flat channel flow are now available up to $Re_\tau=5200$, but simulations of flat plate turbulent boundary layer are more challenging and therefore restricted to slightly lower Reynolds numbers. The behavior of very large scale structures is clearly dependent on the Reynolds number, and they are likely to become significant at extremely high Reynolds number. However, these Reynolds numbers are not accessible by DNS yet. Additionally, statistics of very large scales can be affected by the size of the computational domain or the forcing mechanism of the turbulent boundary layer simulations. Still, DNS is an important tool to study space and time organization of these structures at moderate Reynolds numbers when conducted carefully. Several DNS of TBL have been performed at moderate Reynolds number with various boundary conditions and domain sizes \cite{sillero2013,schlatter2010b,diaz_daniel17}. Higher Reynolds numbers have been reached using wall resolved large eddy simulations (WRLES) i.e., \cite{deck2014,eitelAmor2014}, but the spatial resolution was not enough to investigate the coherent structures down to the buffer region.

The present work includes analyses of the large-scale structures from a new DNS of the flat plate boundary layer at moderate Reynolds number. At this Reynolds number ($Re_{\theta} \simeq 2000$), large-scale structures are already present even if the log region is still not well defined. Recently, Srinath \etal~\cite{srinath2017} proposed a new simple model of streamwise velocity large-scale structures and related this model to the $k_x^q$ (with $q \simeq -1$) streamwise energy spectra in the low wavenumber range. This model relies on a given length distribution of the large-scale structures of the fluctuating streamwise velocity.

The article is organized as follows. In Section \ref{sec:numerics}, turbulence statistics, energy spectra and two-point spatial correlations are discussed following a brief description of the numerical code and the details of the new databases. The detection method used to extract the large-scale motions is described, and statistical results are analyzed in Section \ref{sec:lsm}. Conclusions are drawn in Section \ref{sec:conc}.

\section{\label{sec:numerics} Numerical simulation}

\renewcommand{\thefootnote}{\fnsymbol{footnote}}

The DNS of TBL was performed using the code Incompact3d\footnote[2]{\url{https://www.incompact3d.com/}}.
The code is massively parallel and solves the incompressible Navier-Stokes equations. It uses sixth order compact finite difference schemes for the spatial discretization and different time schemes are implemented (Adams-Bashforth or Runge-Kutta). Inflow/outflow, periodic, free-slip, no-slip boundary conditions are available. Fractional step method ensures incompressibility condition which requires the solution of a Poisson equation for the pressure. The Poisson equation is solved in spectral space using Fast Fourier Transforms. Laizet and Lamballais \cite{laizet2009} proposed a method to use non-homogeneous grid spacing in one direction based on a single analytic stretching function to keep the benefit of the accurate solution of the spectral pressure treatment. Combined with the concept of the modified wave number, this direct technique satisfies the divergence-free condition up to machine accuracy while also supporting the use of a stretched mesh in one direction. A partially staggered mesh is used for the pressure to avoid oscillations.

In the present study, the code was used to simulate a turbulent boundary layer with a domain of size $L_x=600\delta_{o}$,  $L_y=40\delta_{o}$,  $L_z=20\delta_{o}$ where  $\delta_{o}$ is the boundary layer thickness of the inlet Blasius profile and $x,y,z$ are the streamwise, wall-normal and spanwise directions. Advection condition is used at the outlet of the simulation domain for all velocity components using the local streamwise value $u_c = u(L_x, y, z, t)$ while backflow is avoided with the additional condition to eliminate any possible negative convection velocity. No-slip and homogeneous Neumann boundary conditions are used for the bottom wall and the upper boundary respectively. The periodic boundary conditions used in the spanwise direction are implemented into the solver by the use of appropriate compact finite difference schemes.

\begin{figure}
    \centering
    \includegraphics{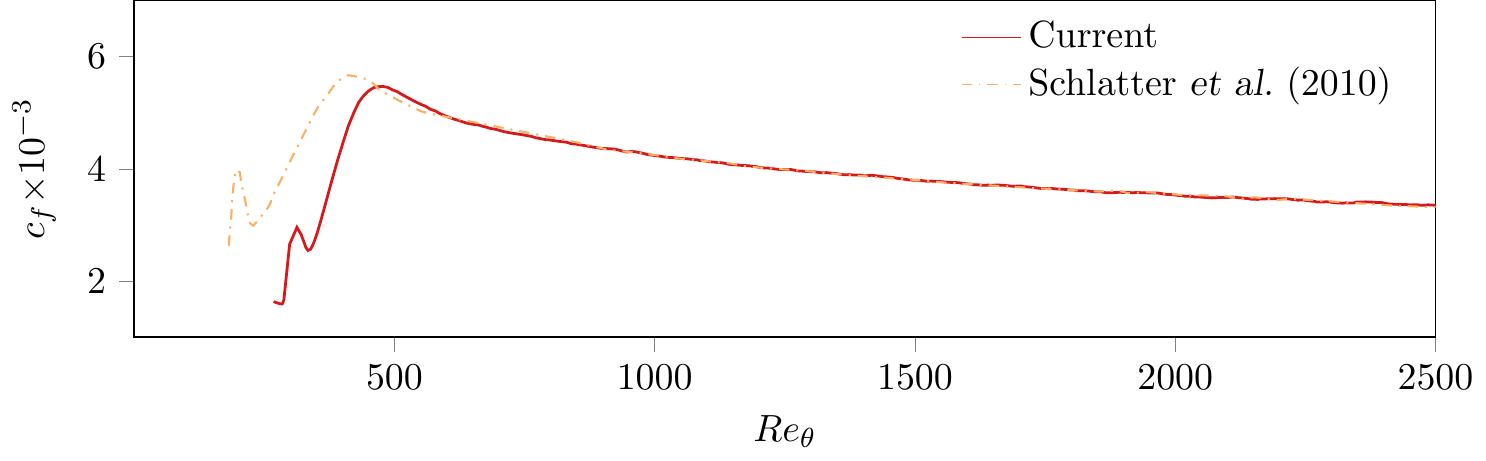}%
    \caption{Comparison of the friction coefficient $c_f$ with the DNS of Schlatter \etal~\cite{schlatter2010b}}
    \label{fig:cf}
\end{figure}

Different methods have been proposed to generate turbulent inlet conditions for DNS of turbulent boundary layers. The recirculation introduced by Lund \etal~\cite{lund1998} can be used to limit the extent of the simulation domain for very high Reynolds numbers. However, the periodicity when associated with a too short simulation domain can affect the statistics of the very long structures. A most suitable solution is to start with a laminar boundary layer profile and to force the transition with a tripping function. Schlatter and Orlu \cite{schlatter2012b} investigated the effects of the different transition procedure on the flow statistics and provided an efficient tripping mechanism. Following their work, a random source term is added to the momentum equation in wall-normal direction within a limited volume close to the bottom wall. The effect of the equation is similar to a sandpaper strip used in experimental studies. It is applied at $14 \delta_o$ away from the inlet, close to the wall with an extent of $l_x= 1.4 \delta_o$ , $l_y= 0.35 \delta_o$ in the streamwise and wall-normal directions while covering the entire spanwise direction. The Reynolds number at the tripping position is $Re_\theta \simeq 300$ and the temporal and spanwise cutoff scale of the tripping are set to $t_s = 1.4 \delta_o / U_\infty$ and $z_s = 0.6 \delta_o$ respectively. The amplitude is tuned to minimize overshooting in the resulting friction coefficient (Figure~\ref{fig:cf}).

The stretching parameters were adapted to keep the resolution in the wall-normal direction below $10$ wall units over most of the turbulent region. The grid spacings in the two other directions are in agreement with the usual practice for the DNS of such flows ($\Delta x^+ \simeq 8$ and $\Delta z^+ \simeq 4$). The simulation parameters are summarized in Table~\ref{tab:param}.

\begin{table}[h]
    \caption{Parameters of the turbulent boundary layer at $Re_\theta = 2068$ for which the boundary layer thickness $\delta$ is equal to $8.46 \delta_o$. $L_{x}$, $L_{y}$ and $L_{z}$ are the sizes of the simulation domain, $N_{x}$,\,$N_{y},\, \textrm{and}\, N_{z}$ the corresponding resolution. The reference momentum thickness $\theta$ and the friction velocity $u_\tau$ are also taken at the same streamwise position.}
\setlength{\tabcolsep}{12pt}
    \begin{center}
        \begin{tabular}{@{}cccc}\toprule
            $Re_{\theta}$ & $\left(L_{x},\,L_{y},\,L_{z}\right)/\delta$ & $\Delta x^{+} ,\, \Delta y^{+}_{min} ,\, \Delta z^{+}$  &  $N_{x}\,\times\,N_{y}\,\times\,N_{z}$ \\
            \midrule
            $250-2500$  &  $53.19,\, 4.72 ,\, 2.36$ &  $8.27 ,\, 1.0 ,\, 3.94$ & $6401\,\times\,321\,\times\,448$ \\
            \bottomrule
        \end{tabular}
    \end{center}
    \label{tab:param}
\end{table}

The simulation was performed on 2048 cores of Intel Xeon E5-2690v3. Two separate datasets were collected to converge the statistics on large-scale structures. The first dataset consists of 3D velocity and pressure fields, collected every 500-time steps ($0.0168\, \delta/u_\tau$ based on outlet quantities) for the full simulation domain. For the second dataset, the same quantities were recorded at 4 spanwise-wall normal planes ($Re_\theta=922,\,1522,\,2068$ and $2365$) for every 5-time steps corresponding to a streamwise displacement of half a grid spacing based on the free stream velocity. Details of the time-resolved databases are summarized in Table~\ref{tab:database}. The time-resolved datasets will be used for the validation of Taylor's hypothesis and the comparison of spatial-temporal features of the investigated turbulent structures. These last datasets can be used as turbulent inlet condition for future simulations of the turbulent boundary layer (e.g., TBL with adverse pressure gradient) as the extent of the time-resolved data is enough to conduct new simulations over multiple characteristic times.

\begin{table}[h]
        \caption{Parameters of the 2D (spanwise - wall-normal plane) time-resolved database. The boundary layer thickness $\delta$ and the friction velocity $u_\tau$ are evaluated at the streamwise location $x$ given in fractions of the full domain size $L_x$. Statistics are calculated over the total time $T$.}
\setlength{\tabcolsep}{12pt}
    \begin{center}
        \begin{tabular}{@{}ccccc}\toprule
            $x$ & $Re_{\theta}$ & $Re_{\tau}$ & 
            $\delta/\delta_{o}$ & 
            $Tu_{\tau}/\delta$\tabularnewline
            \midrule
            $0.25L_x$ & $ 924$ & $374$ & $3.98$ & $33.4$ \\
            $0.50L_x$ & $1527$ & $552$ & $6.26$ & $20.0$ \\ 
            $0.75L_x$ & $2068$ & $722$ & $8.46$ & $14.3$ \\ 
            $0.90L_x$ & $2371$ & $813$ & $9.58$ & $12.5$ \\
            \midrule
        \end{tabular}
    \end{center}
    \label{tab:database}
\end{table}

\subsection{\label{subsec:turb_stats_spec}Turbulence statistics \& energy spectra}

\begin{figure}
    \centering
    \begin{minipage}[b]{0.99\columnwidth}
            \includegraphics{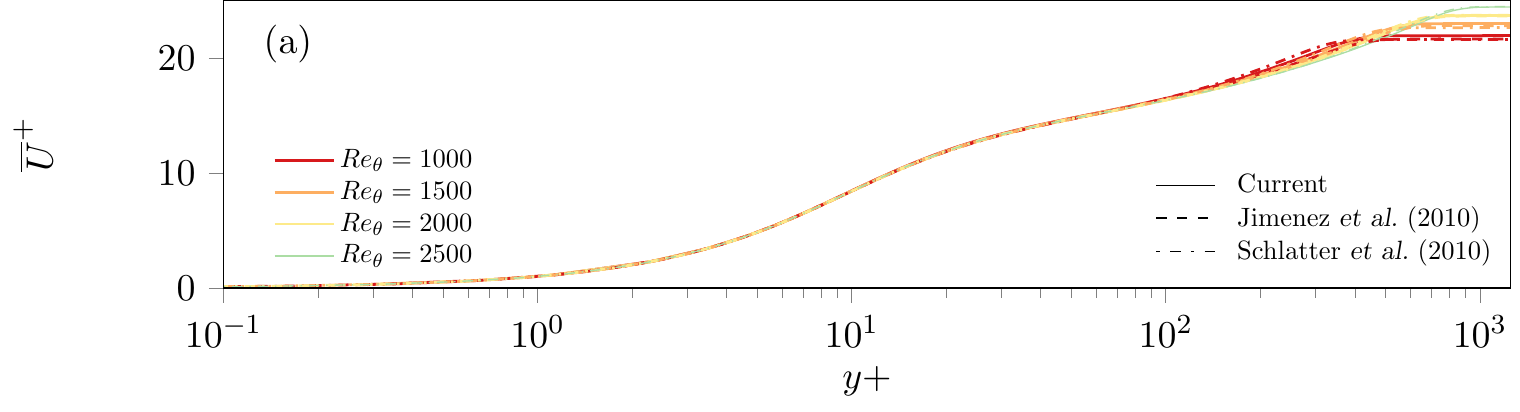}
        \end{minipage}

        \begin{minipage}[b]{0.99\columnwidth}
            \includegraphics{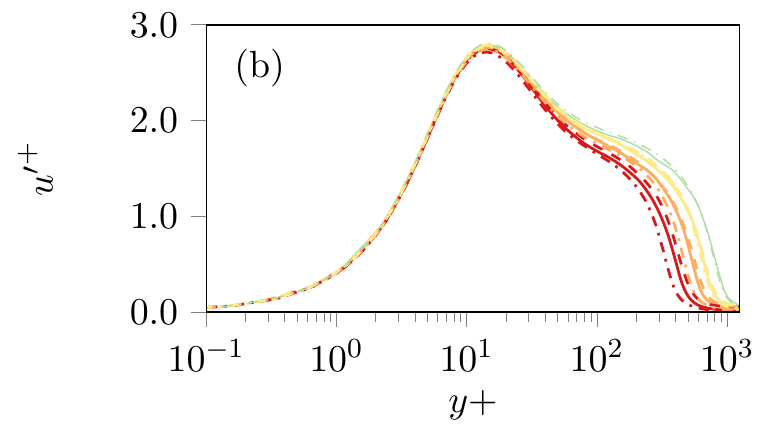} 
            \includegraphics{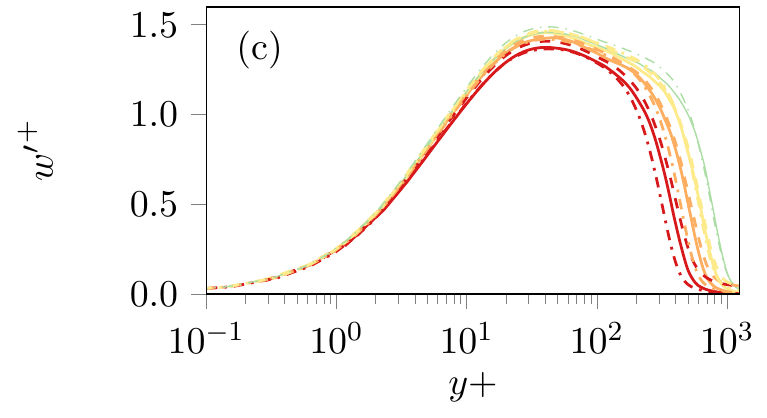}
        \end{minipage}

        \begin{minipage}[b]{0.99\columnwidth}
            \includegraphics{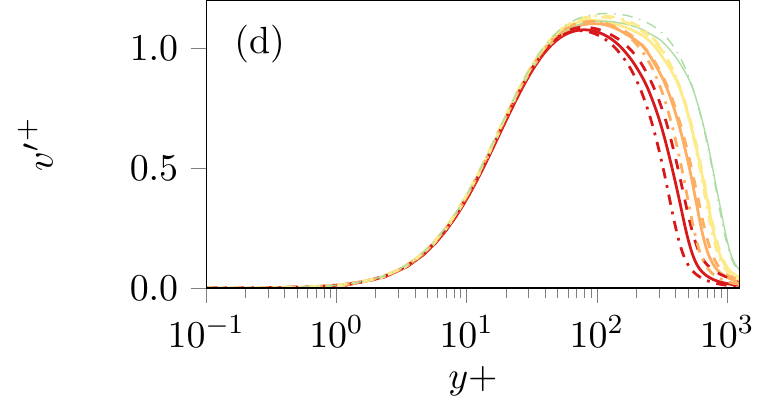}              
            \includegraphics{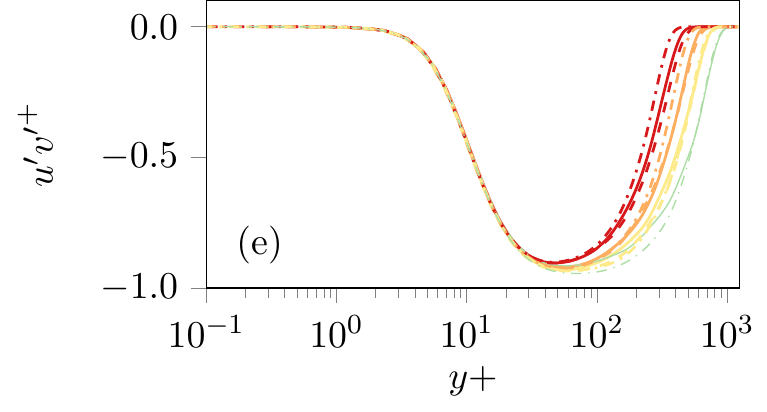}
        \end{minipage}

        \caption{\label{fig:stat} Mean velocity and Reynolds stresses (continuous) profiles of the current study in comparison with profiles from (dashed) Jimenez \etal~\cite{jimenez2010} (at $Re_\theta=1100,\,1551, 1968$)  and (dash dot) Schlatter \etal~\cite{schlatter2010b} at $Re_\theta=1000,\,1410,\,2000, \textrm{ and } 2540$.}
\end{figure}

To validate the results of the present simulation mean profiles of the streamwise velocity and the turbulent intensities are compared at $3$ Reynolds numbers with the DNS of Jimenez \etal~\cite{jimenez2010} and Schlatter \etal~\cite{schlatter2010b} (see Figure~\ref{fig:stat}). The slight differences may be attributed to the different spatial resolutions (especially in the outer part of the TBL) as well as different inlet conditions. 

\begin{figure}
  \centering
  \begin{minipage}[b]{0.495\columnwidth}
        \centering
        \includegraphics{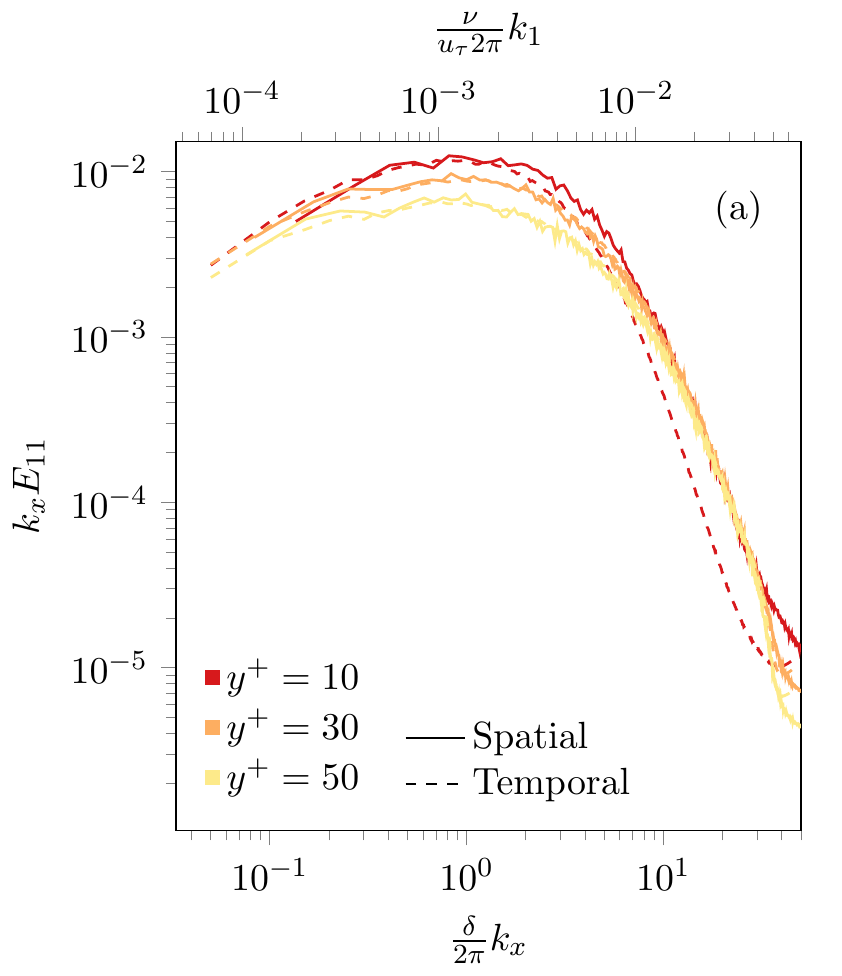}
  \end{minipage}
  \begin{minipage}[b]{0.495\columnwidth}
        \centering
        \includegraphics{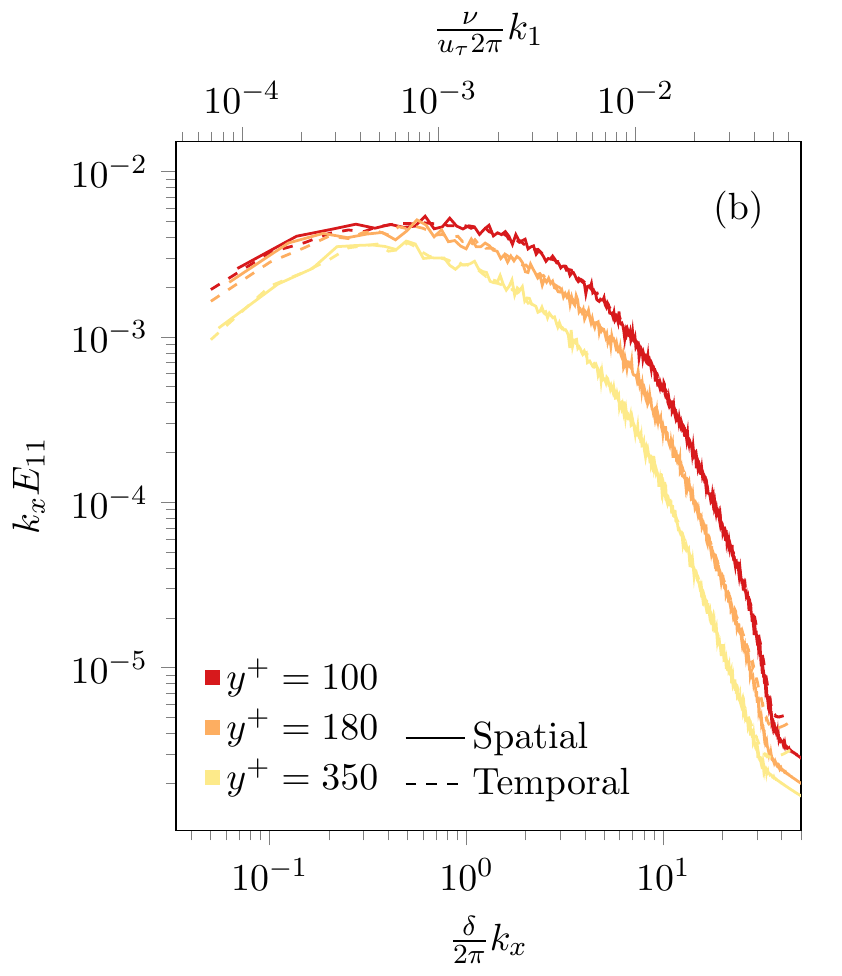}
  \end{minipage}

    \caption{Streamwise energy spectra using spatial (continuous) and temporal (dashed) data at different wall distances. The time spectra are computed at the streamwise position such that $Re_\theta=2068$ using the Taylor hypothesis and the spatial one are computed in a domain such as $1764<Re_\theta<2348$ which corresponds to approximately $20$ local boundary layer thickness at $Re_\theta=2068$.}
    \label{fig:spectra_all}
\end{figure}

As both time and space spectra are available at $Re_\theta = 2068$, the validity of Taylor's hypothesis can be evaluated very close to the wall. 
The two types of spectra are compared in Figure~\ref{fig:spectra_all} using mean velocity profile as convection velocity. They are almost identical above $y^+=50$ but start to depart from each other at $y^+=30$ indicating the limit of Taylor's hypothesis with this choice of convection velocity.
It is well known that in the flow regions where shear is dominant (like the near wall region of TBL) Taylor's hypothesis does not hold since local mean velocity is not equal to the convection velocity. Therefore, the use of Taylor's hypothesis with this convection velocity is not suitable for the detection of the structures down to the bottom of the buffer region. 
In the current study, this limit will be used to define attached structures from spatial data, i.e., all the structures starting $y^+=50$ or below.
As a consequence, our detection criteria for these attached structures should not be altered by the Taylor hypothesis. 

In the framework of the Townsend Eddy model, the attached eddies are associated to a $k_x^{-1}$ streamwise energy spectra close to the wall on a limited range of wall distances at sufficiently large Reynolds numbers. 
Baars \etal \cite{baars2017} identified universal attached eddies for relatively high Reynolds numbers from careful interpretation of coherence spectrogram while emphasizing the need for the unobstructed view of a $k_x^{-1}$. 
 As mentioned, Srinath \etal~\cite{srinath2017} have shown that, even at moderate Reynolds number, a $k_x^{-q}$ slope can be observed in the buffer and mesolayer with increasing value of $q$ such as $q=1$ is valid only at a specific wall distance between 100 and 200 wall units at the investigated Reynolds numbers. This $k^{-1}$ region was already observed by George \& Tutkun \cite{george2011} in the outer part of the mesolayer ($y^+ \sim 200$) from WALLTURB hot-wire data at the same Reynolds number as \cite{srinath2017}. The streamwise velocity fluctuations spectra of the present DNS at $Re_\theta= 2068$ are shown in Figure~\ref{fig:spectra_all} for time-resolved and spatial databases. As the Reynolds number is not very high, the  $k_x^{-q}$ range is not clearly visible. Nevertheless, the streamwise energy spectrum near $y^+=100$ has the most compatible results with a $k_x^{-1}$ scaling. The value of $q$ increases when moving from the wall. However, as the local Reynolds number is $Re_\tau=722$, the logarithmic region of the mean streamwise velocity profile is short, and $y^+=250$ is already located in the wake region.

\subsection{\label{subsec:corr}Two-point spatial correlation}

\begin{figure}

    \includegraphics{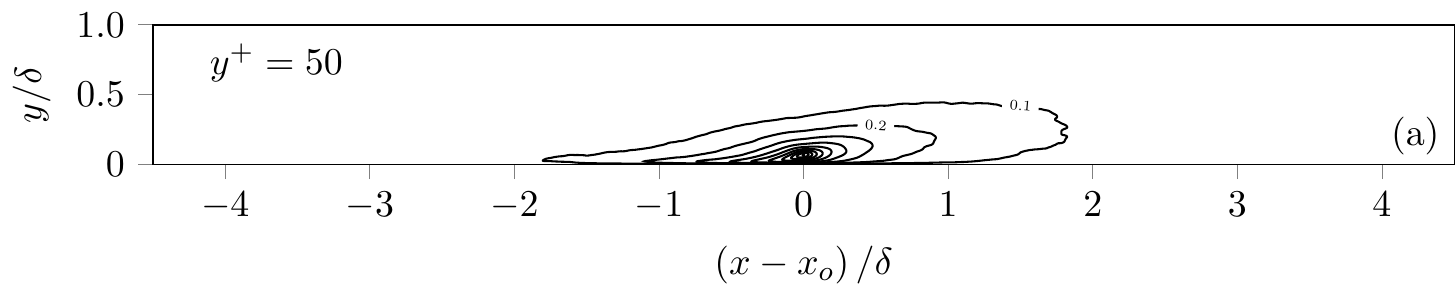}

    \includegraphics{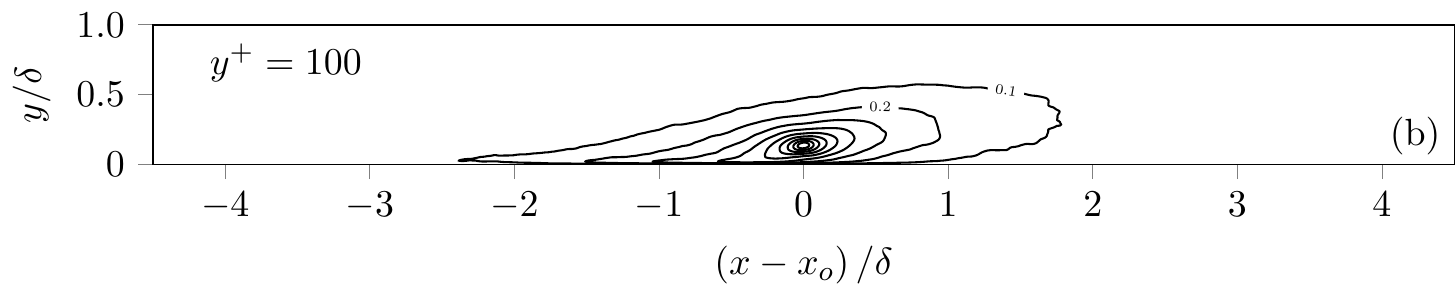}

    \includegraphics{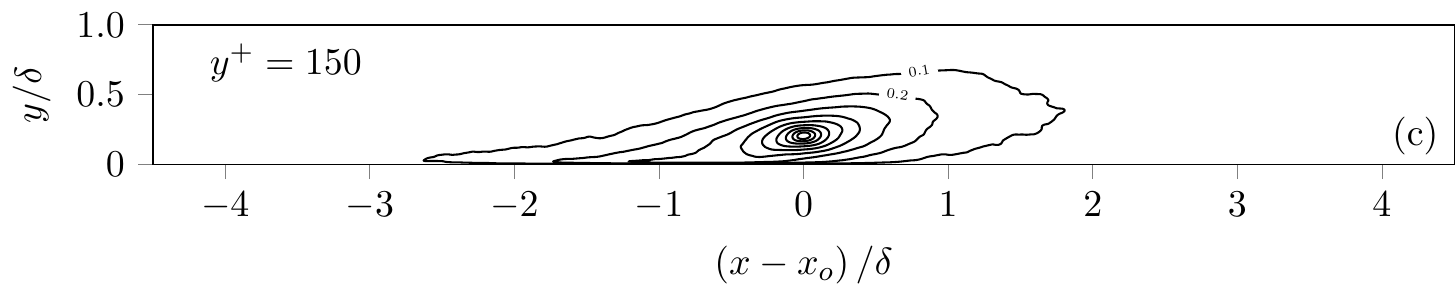}

    \caption{Two-point spatial correlation function of the streamwise velocity fluctuations $\langle u(x{-}x_o,y{-}y_o)u(x,y)\rangle$ at wall distances (a) $y_o=50^+$, (b) $y_o=100^+$ and (c) $y_o=150^+$  where $x_o$ is the streamwise position such that $Re_\theta=2068$}%
    \label{fig:2p_corr}
\end{figure}

The average statistics of the large-scale motions can be found with two-point correlations. Figure~\ref{fig:2p_corr} shows the two-point correlation of the streamwise velocity fluctuations at three wall distances in the streamwise wall-normal plane. Correlation isocontours from $0.1$ to $1$ with a $0.1$ increment are plotted as functions of $(x-x_o)/\delta$ and $y/\delta$ where $x_o$ is the streamwise position for which $Re_\theta=2068$. The spatial correlations extend over a distance of $4 \delta$ for all wall distances based on the $0.1$ correlation isocontours. Tutkun \etal~\cite{tutkun2009} demonstrated similar correlations using data obtained by hot-wire rakes at a much higher Reynolds number ($Re_{\theta} = 19100$). The correlation isocontours exhibit an elliptical shape with an average inclination function of the altitude of the fixed points. Different results have been reported for an average angle ranging from $9^\circ$ to $33^\circ$ \cite{fage1932, corino1969, brown1977, falco1977}. The inclination of the elliptical shape of isocontours varies with the wall distance. The overall streamwise extent of the correlation grows with distance from the wall in the logarithmic region (also broadens in the wall normal direction) but drops beyond that region in agreement with Ganapathisubramani \etal~\cite{ganapathisubramani2005}. As stated before, such correlations reflect the average length, height and eventually width of streamwise structures if performed in 3D, like in Sillero \etal~\cite{sillero2014}. The extends of the low but significant values of the two-point correlations at large separations indicate the average size of the largest structures but it does not show much about the different lengths of the structures.

\section{\label{sec:lsm}Large-scale motions}

\subsection{Detection methods}

The analysis of large-scale structures of streamwise velocity fluctuations is conducted on a domain of 20 local boundary layer thickness centered at $Re_\theta=2068$  ($0.6 L_x < x < 0.9 L_x$) extending up to $Re_\theta=2407$ sufficiently remote from the outlet not to affect the statistics.

Correlation functions are not able to give specific information about the space and time extent of a single structure but rather provide an average view. The large-scale structures can be characterized individually by taking the benefit of the 3D database. Zaki \cite{zaki13} proposed a detection method based on local maxima or minima of the streamwise velocity fluctuations. However, such a method is more adapted to regular streamwise structures. In the present study, we used another method based on simple thresholding of the streamwise velocity fluctuations. The same method was also used to study the time-resolved evolution of coherent structures from a DNS of channel flow \cite{lozanoduran2014b} and to characterize the near wall streaks from PIV databases \cite{lin2008}. Extraction procedure consists of two steps: 
\begin {enumerate*} [label=(\textcolor{blue}{\roman*})]
\item generation and \item labeling of binary images. 
\end {enumerate*} 
No other treatment such as additional filtering or morphological operations is necessary.

Binary images $\mathbb{B^{\ominus}}$ and $\mathbb{B^{\oplus}}$ indicative of negative and positive streamwise fluctuations are obtained respectively by,

\begin{equation}
    \mathbb{B}^{\ominus} = \left\{\!
 \begin{array}{ll}
         1 \;\; \mbox{if} \;\; u^\prime\! <\! C_{thr} \, \sigma_u^{100^+} \\
         0 \;\; \mbox{otherwise} \! \\
     \end{array} \!
     \right. , \;\;\;\;
     \mathbb{B^{\oplus}} = \left\{\!
\begin{array}{l}
         1 \;\; \mbox{if} \;\; u^\prime\! >\! C_{thr} \, \sigma_u^{100^+} \\ 
         0 \;\; \mbox{otherwise} \! \\
     \end{array} \!
     \right. \!
     \label{eq:thr}
\end{equation}    
where $\sigma_u^{100^+}$ is the standard deviation of the streamwise velocity at $y^+ = 100$, and $C_{thr}$ is the threshold parameter. 

A second (outer) peak of streamwise turbulence intensity comes out between the buffer and the log region as the Reynolds number increases. These results have been observed in many independent experiments~\cite{hutchins2007b,hultmark2012,srinath2017}
. The second peak of ${u^\prime}^+$, which is rather a plateau for not sufficiently high Reynolds numbers, is the result of the strengthening large-scale streamwise structures \cite{vallikivi15b}. Even though there is no secondary peak or plateau at the moderate Reynolds number of the present simulation, $\sigma_u^{100^+}$ gives a good estimation of what would be the intensity of this second peak at much larger Reynolds numbers. Therefore this value is taken as a reference intensity for the large-scale structures. An analysis is conducted with three different values of the threshold coefficient $C_{thr}$ (0.5, 1.0 and 1.5) to demonstrate its effect on statistics. 

Percentages of the retained energy, momentum and volume fractions with different thresholds are given in Table~\ref{tab:detec_fraction}. 
Such a detection method retains a small fraction of the volume but a significant fraction of the streamwise turbulent kinetic energy.
A more detailed repartition is given in Figure 5 with the weighted PDF of streamwise velocity fluctuations. 
The distribution of streamwise energy due to positive and negative fluctuations are similar at $y^+=20$ but start to differ when moving from the wall where a sharper distribution is observed for the contribution of positive fluctuations. 
Consequently, the energy contribution of the detected high-speed velocity structures is much lower than the low-speed ones and become very small above $y^+=450$ ($\sim 0.6\delta$) with $C_{thr}=1$.

\begin{table}[h]
    \caption{Streamwise energy, momentum and volume fraction (in \% of the total) after thresholding only, low momentum regions (${\ominus}$)  and high momentum regions (${\oplus}$) extracted using Eq. (\ref{eq:thr}) as function of the threshold parameter $C_{thr}$. }
    \begin{center}
        \begin{tabular}{@{}ccccccccccccc}
            \toprule
            & &\multicolumn{3}{c}{Energy} & & \multicolumn{3}{c}{Momentum} & & \multicolumn{3}{c}{Volume}
            \tabularnewline
            $C_{thr}$  & & ${\ominus}$  & ${\oplus}$ & $\ominus \cup \oplus$ & &  ${\ominus}$ &  ${\oplus}$ &  $\ominus \cup \oplus$  & &  ${\ominus}$ & ${\oplus}$ &  $\ominus \cup \oplus$\\
            \midrule
            $0.5$   &    &$48\%$&$49\%$&$97\%$&   &    $43\%$&$46\%$&$89\%$ & & $29\%$&$34\%$&$63\%$\\
            $1.0$   &    &$39\%$&$41\%$&$80\%$&   &    $29\%$&$32\%$&$61\%$ & & $15\%$&$17\%$&$32\%$\\
            $1.5$   &    &$25\%$&$25\%$&$50\%$&   &    $15\%$&$16\%$&$31\%$ & &  $6\%$& $7\%$&$13\%$\\
            \bottomrule
        \end{tabular}   
    \end{center}
    \label{tab:detec_fraction}
\end{table}

\begin{figure}
  \begin{center}
    \includegraphics{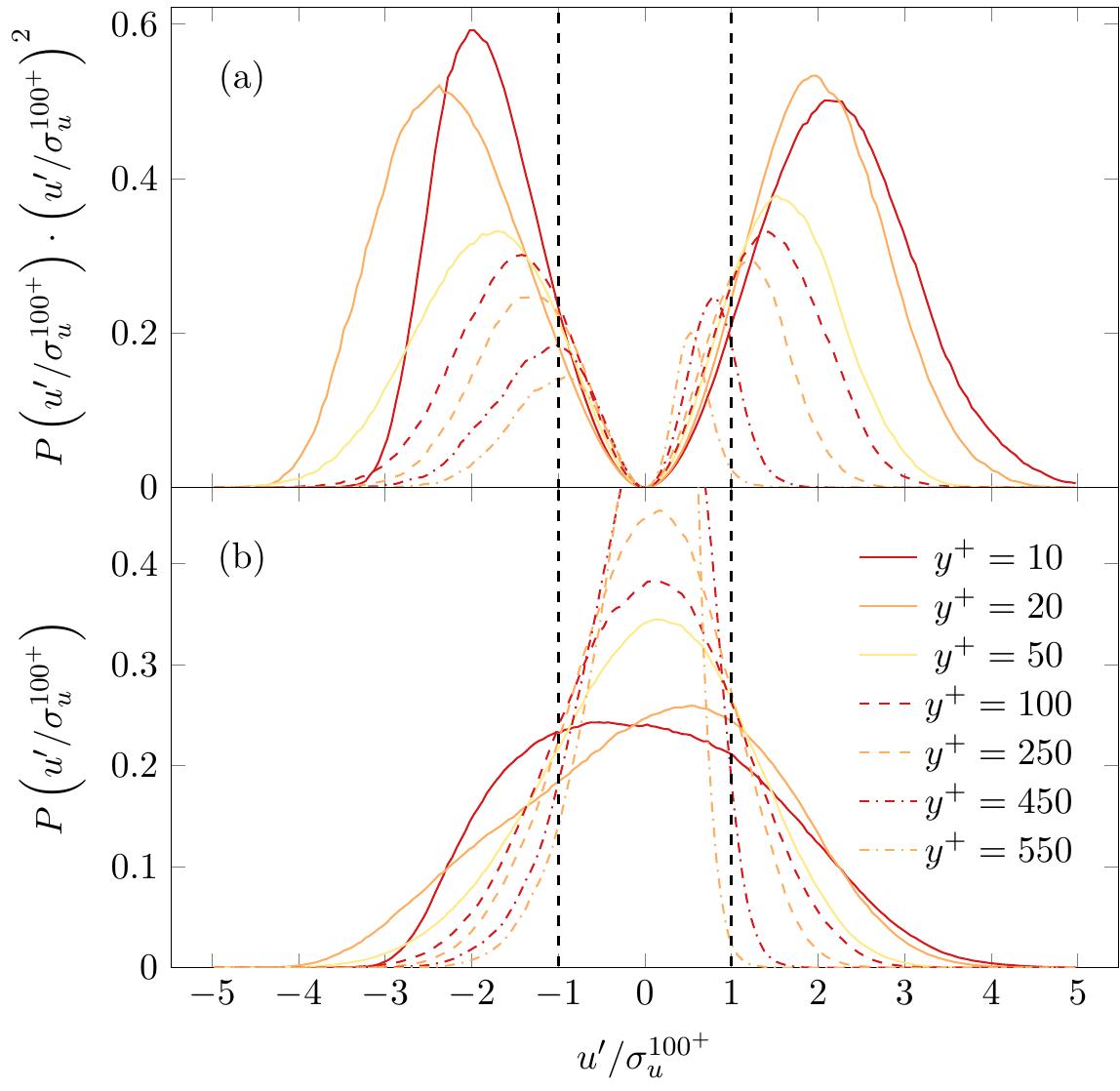}
      \caption{Energy and volume distribution of the streamwise velocity fluctuations. Vertical dashed lines represent the threshold coefficient $C_{thr}=1$ meaning that outer left and outer right regions are kept for low and high momentum regions respectively.}
    \label{fig:c_thr_contr}
  \end{center}
\end{figure}

In a second step, binary volumes ($\mathbb{B^\ominus}$ or $\mathbb{B^\oplus}$) are subjected to the labeling procedure which assigns an index for each individual structure (with no pixel connection with other structures). The detection procedure generates a large number of small structures which are not the primary interest in this study. Therefore, structures with a total length of less than 0.2 local boundary layer thickness in the streamwise direction ($\lambda_x$) are discarded. This procedure is not essential but reduces the total number of structures significantly to be analyzed without affecting the statistics of the large ones. This study focuses on the attached structures but the detection procedure based on a threshold of the fluctuating streamwise velocity makes it impossible for a detected structure to touch the wall as fluctuations go to zero. Hence all structures with a minimum wall distance larger than $50^+$ are discarded.

Extracting a single scale for complex multi-branch structures is probably not meaningful. Therefore, to better characterize their complexity, the skeletons of the structures are determined. This method simplifies the 3D binary volume of a single structure (individual objects extracted from $\mathbb{B^{\ominus}}$ and $\mathbb{B^{\oplus}}$) to a set of curves using a skeletonization algorithm. The advantage of this simplification is that quantitative statistics of each branch of the skeleton can be extracted. The simplest method, known as  ``thinning'', can provide skeletons but the results are very sensitive to the surface smoothness of the analyzed volume. In the present study, this method would lead to very complex skeleton topologies which need to be simplified to gather useful statistics. The thinning has been applied successfully for instance by Marquillie \etal~\cite{marquillie11} for the detection of the near wall streaks which are more regular and therefore easier to characterize.

The method which has been used in the present analyses is a more robust algorithm for computing continuous sub-voxel accurate curves from volumetric objects. The basis of the method was developed by Hassouna \etal~
\cite{hassouna2009} and has been adjusted for this work. It can represent a complex structure with a limited number of curves. Unlike other proposed methods to extract turbulent structures, the curves of the skeletons are not necessarily associated with the local maxima of the quantity to analyze (e.g., streamwise fluctuations). The skeleton only represents the global shape and the geometric complexity of the volumetric structures. 
One parameter of the method controls the degree of refinement of the skeletons independently from the surface property of the associated volume. 
This feature also comes with reliable robustness, as the shape of the detected curves depends only weakly on the detail of the structure unlike other methods based on the scalar value or other skeletonization methods like thinning which leads to very complex skeletons highly dependent on the smoothness of the surface of the structure.
The method requires significant computational resources for the extraction of the well-resolved very large structures as several eikonal equations have to be solved for each detected structure. Details of the method are given in Appendix \ref{app:skel}.

The skeletonization procedure has been applied to the same 20 local boundary layer thickness long 3D sub-domains centered at $Re_\theta=2068$. Skeletons are extracted from structures in the binary volume $\mathbb{B^{\ominus}}$ or $\mathbb{B^{\oplus}}$ after interpolation of the velocity fields on a regular isotropic grid with a $6.7^+$ mesh size as the skeletonization procedure used in this study requires an isotropic discretization. Moreover, a weak cleaning procedure composed of opening and closing operations is applied to the binary objects before applying the skeletonization. The cleaning steps are required to better resolve the small connections between selected regions initially determined by one or two pixels only. A minimum resolution of 3 to 5 pixels is needed to be able to extract a skeleton. When the soft cleaning procedure is not enough, the binary volume is interpolated on a finer grid to resolve sufficiently the bottleneck regions. 
Skeletons are extracted only for the large structures to concentrate the analysis on the largest structures.

\begin{figure}
        \begin{center}
        \includegraphics{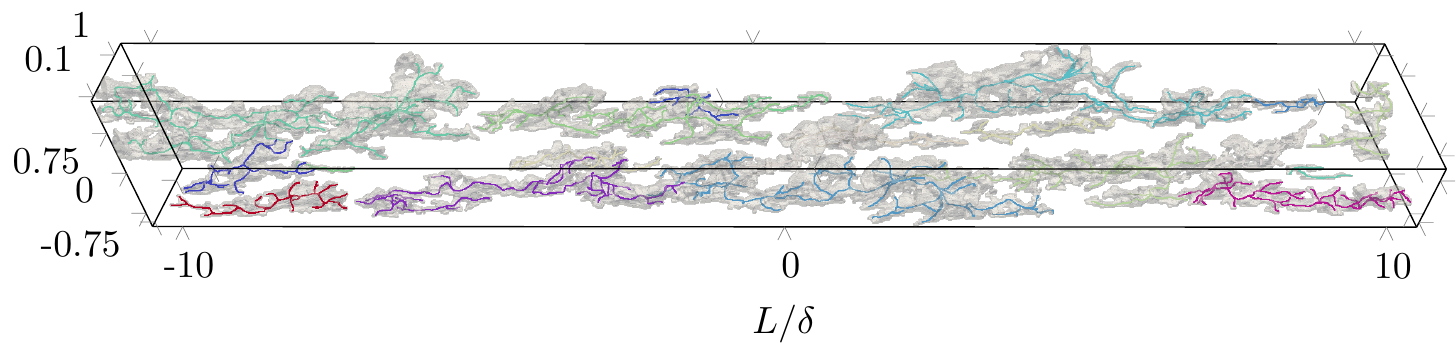}
            \caption{\label{fig:skeleton_3d_all} Isovolume of the 3D streamwise velocity fluctuation structures with their skeletons extracted using thresholding (Eq. (\ref{eq:thr})) with $C_{thr}=1.0$ on a $20 \delta$ long sub-domain centered at the Reynolds number $Re_{\theta}=2068$. Different colors are associated to separated structures.}
        \end{center}
    \end{figure}

\begin{figure}
        \centering

            \includegraphics{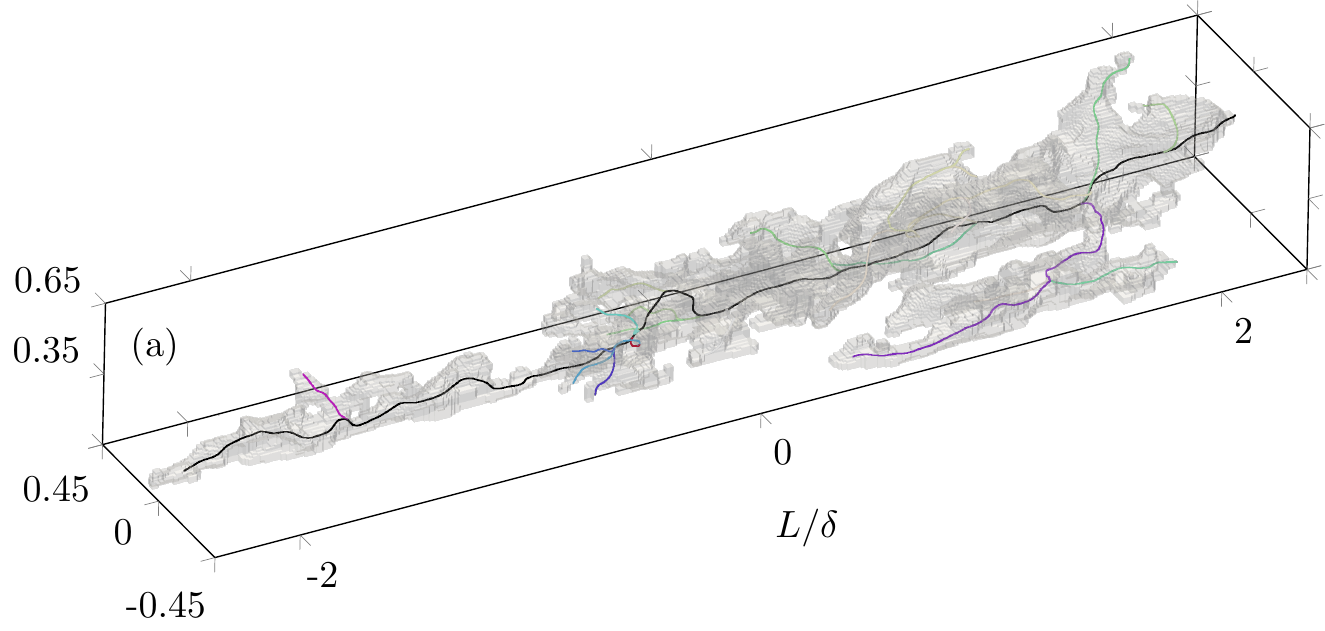}

            \includegraphics{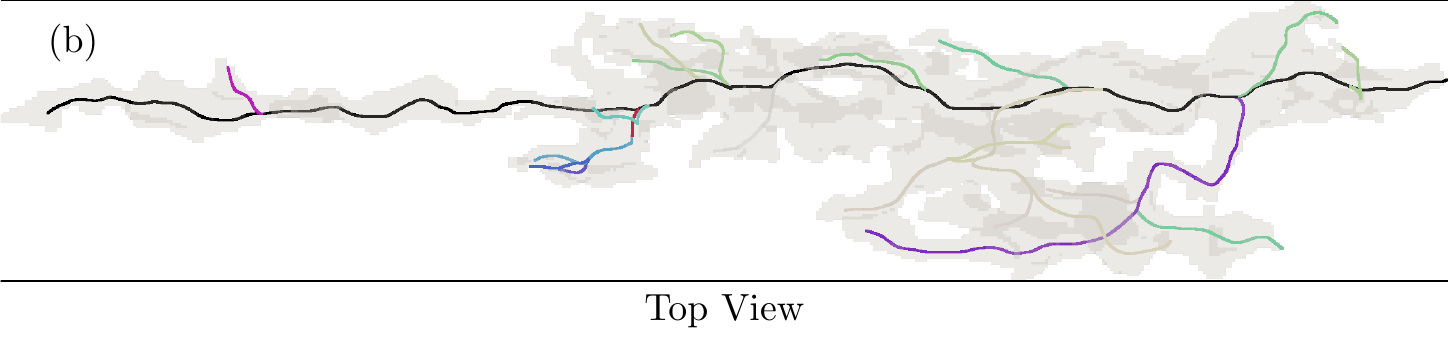}

            \includegraphics{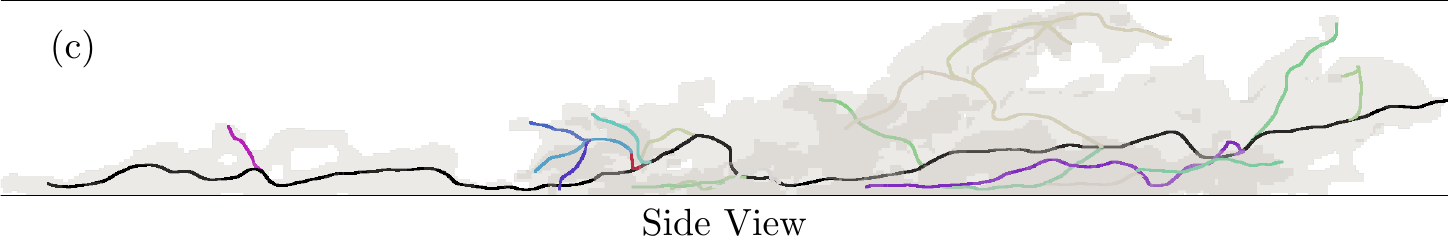}

        \caption{\label{fig:skeleton_3d_single} Skeleton of a single structure. The black curve represents the longest curve (referred as the main curve), and colorful curves are the branches.(a) Isometric, (b) top and (c) side views of the geometry.}

    \end{figure}

The result of the skeletonization in a sub-domain is shown in Figure~\ref{fig:skeleton_3d_all}. The procedure is able to reflect the complex multi-branched behavior of the detected binary volumes with 1D sub-pixel lines for which we can extract a topology, unlike the classical skeletonization algorithms as thinning which only provides pixelized information. The complex shape of the skeleton is presented for a single structure in Figure~\ref{fig:skeleton_3d_single}. The curves of the skeletons reveal the meandering nature of the structure in both the spanwise and the wall normal directions. A side-view of the single structure with its skeleton shows the possible complexity of their shape and the difficulty of defining a single mean angle of a structure.

\subsection{Results}

\begin{figure}
  \begin{minipage}[b]{0.495\columnwidth}
        \centering
        \includegraphics{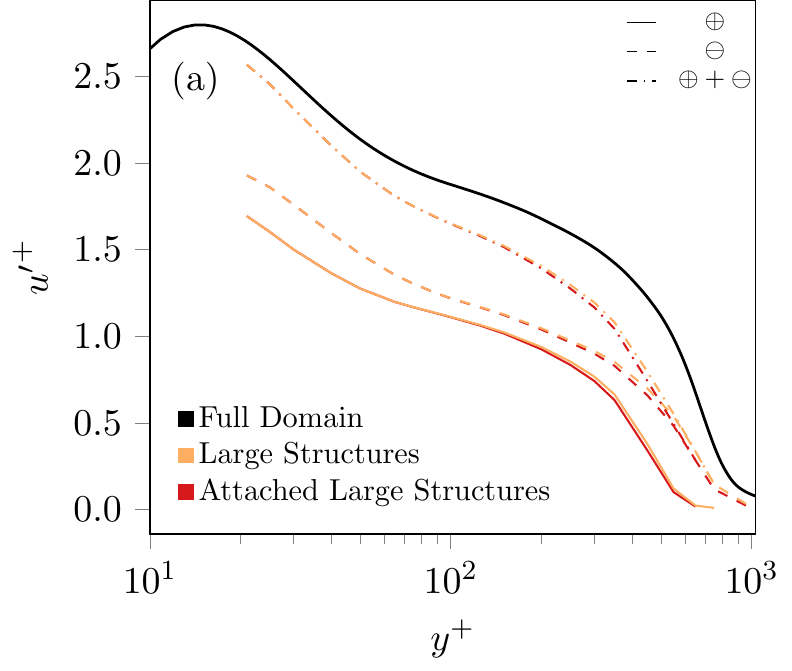}
  \end{minipage}
  \begin{minipage}[b]{0.495\columnwidth}
        \centering
        \includegraphics{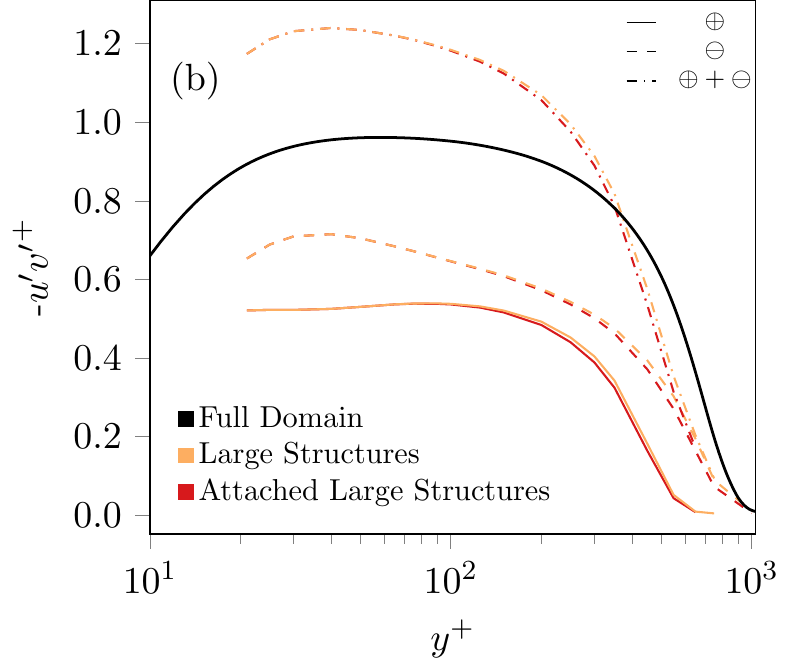}
  \end{minipage}

    \caption{Profiles of streamwise fluctuation (a) and Reynolds shear stress (b) computed from the detected structures with $C_{thr} = 1.0$. Note that, in this figure only, structures not completely included in the  $20\delta$ long investigated domain are kept. Statistics are given for $y^{+} > 20$ which corresponds to the lower bound of the domain to detect structures.}
    \label{fig:energy_structures} 
\end{figure}

The mean profiles of streamwise fluctuations inside the structures are compared to the same quantity for the whole domain to characterize the turbulence statistics of the detected structures (Fig.~\ref{fig:energy_structures}).
The profiles for both positive and negative velocity structures exhibit a similar shape than the profile for the whole domain. 
The streamwise turbulent kinetic energy inside detected structures after cleaning the smallest structures represents approximately 72\% of the same quantity for the whole domain suggesting that the contribution of the detached structures is small. 
Due to our definition of attached structures (with a minimum distance from the wall smaller than $50^+$), the profiles including and excluding the detached structures are identical up to $50^+$. 
However, it is shown that the detached structures do not contribute significantly even in the outer region.
Low-speed detected structures contains more energy than the high-speed ones in agreement with the results of Figure~\ref{fig:c_thr_contr}. 
As already noticed, the energy contribution of positive detected structures drops near $y \sim 0.6\delta$.
A similar analysis of the content of detected streamwise velocity structures was proposed by Ganapathisubramani \etal~\cite{ganapathisubramani2009}. However, in their study, the structures were classified by the energy spectrum. The contribution of the streamwise velocity fluctuation structures to the total Reynolds shear stress, $\overline{u'v'}$, is also significant ($\sim 72\%$) which indicates that attached structures of streamwise velocity fluctuation dominate the Reynolds shear stress as well as.

Once the structures have been detected, various analyses have been used to characterize them. At first, the edges of the smallest cuboid that contains a structure are measured (respectively $\lambda_x, \lambda_y, \lambda_z$). Additionally, the detection procedure was also performed in 2D for each streamwise-normal (XY) plane of the 3D fields in order to compare the results with a similar method than the one used by Srinath \etal~\cite{srinath2017} on PIV data. The 2D detection procedure does not capture the complexity of the structure in the spanwise direction, and the two methods could lead to some different statistics of the streamwise length as demonstrated by Soria \etal~\cite{soria2016} for the quadrants analyses. 

\begin{figure}
  \centering
  \begin{minipage}[b]{0.495\columnwidth}
      \includegraphics{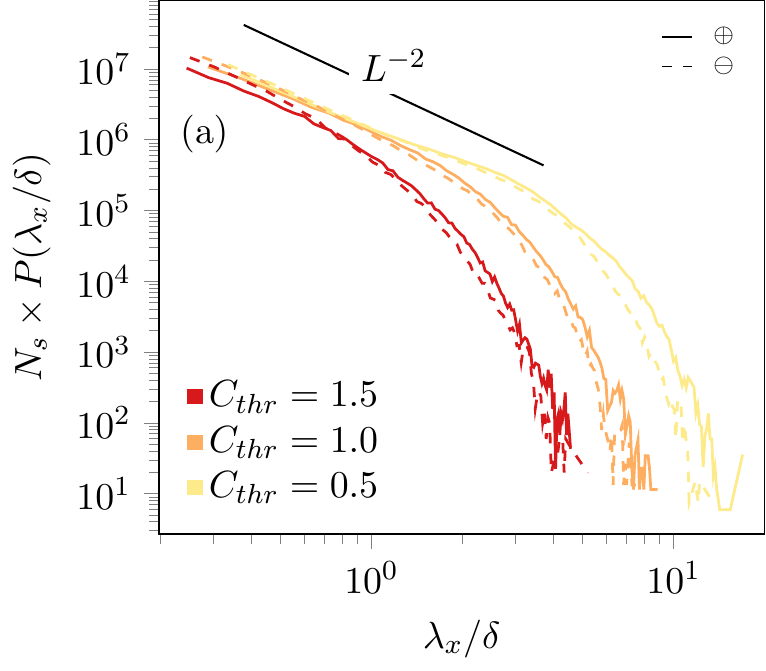}
  \end{minipage}
  \begin{minipage}[b]{0.495\columnwidth}
      \includegraphics{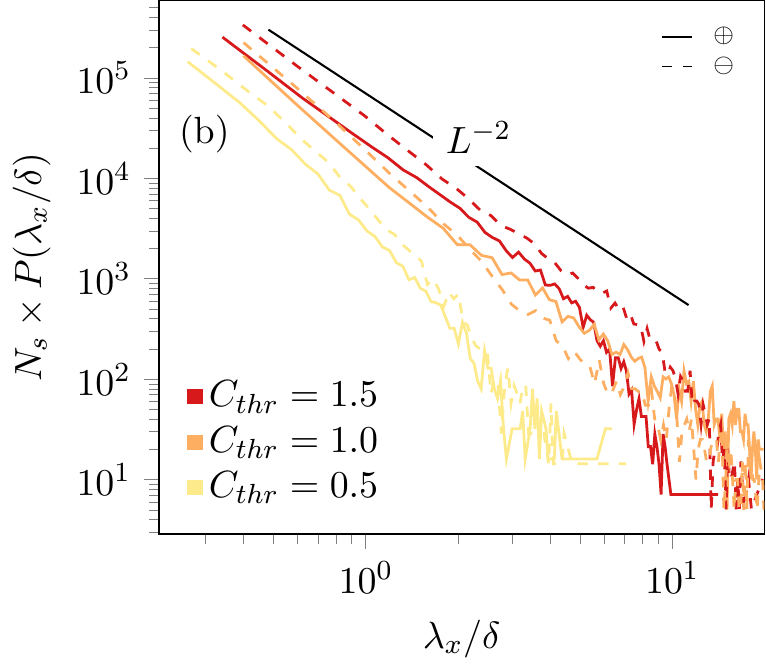}
  \end{minipage}

    \caption{\label{fig:pdf_lambda_x} PDF of the streamwise lengths of the detected structures from positive velocity fluctuations (continuous lines) and negative velocity fluctuations (dashed lines) for three different values of the detection threshold. Statistics are results of (a) 2D detection in XY planes and (b) 3D detection.
    }
\end{figure}

As mentioned earlier, the detection method was applied to a 3D sub-domain of 20 local boundary layer thickness long centered at $Re_\theta=2068$. The probability density functions (PDFs) of the streamwise length premultiplied by the total number of detected structures $N_s$ are shown in Figure~\ref{fig:pdf_lambda_x} for the two different detection procedures and the three values of the threshold coefficient $C_{thr}$ given in Table~\ref{tab:detec_fraction}. The detection of the structures in XY-planes leads to a length distribution $\lambda_x^{-2}$ for the three tested thresholds. This result is in agreement with the results of Srinath \etal~\cite{srinath2017} extracted from PIV at higher Reynolds numbers. 

The comparison of the 2D and 3D detections shows that considering the most intense structures, the two methods lead to a similar length distribution even if their topology is more complex than simple elongated structures. Note that for the 2D detection, the number of large structures decreases when increasing $C_{thr}$. Intuitively, increasing threshold results in shorter structures. However, 3D detection results show the opposite. The main reason is that the largest structures are likely to be connected by the side (in the spanwise direction). As the connections in the spanwise direction are taken into account in  3D detection procedure, when the threshold is too low ($C_{thr}=0.5$), the detection leads to one or few very large structures that cover most of the investigated volume. These structures are larger than the domain of investigation, so they are not taken into account in the statistics of length. This effect explains the smaller number of large structures for $C_{thr}=0.5$  as compared to $C_{thr}=1.0$ for which the largest structures start to be disconnected from each other. 

Unless the results obtained with $C_{thr}$ are biased by the fact that a single long structure usually fills the domain, the probability of length for the different thresholds are qualitatively similar. Therefore, the rest of the analysis will be conducted with $C_{thr}=1$.

\begin{figure}
    \begin{minipage}{0.45\columnwidth}
        \includegraphics{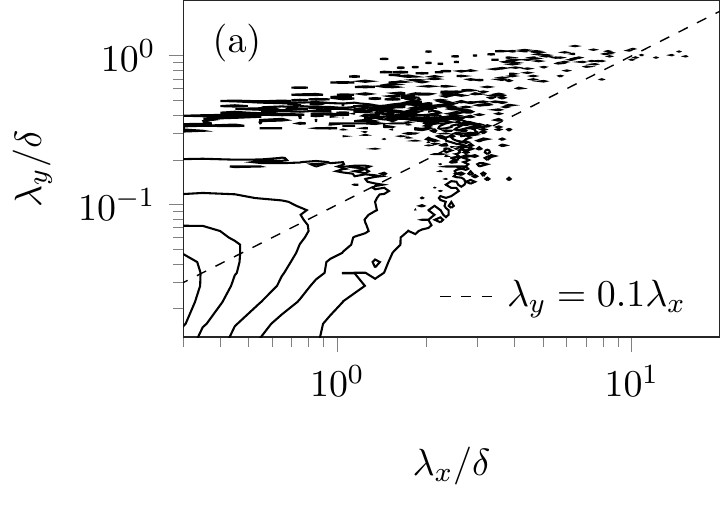}
    \end{minipage}
    \begin{minipage}{0.45\columnwidth}
        \includegraphics{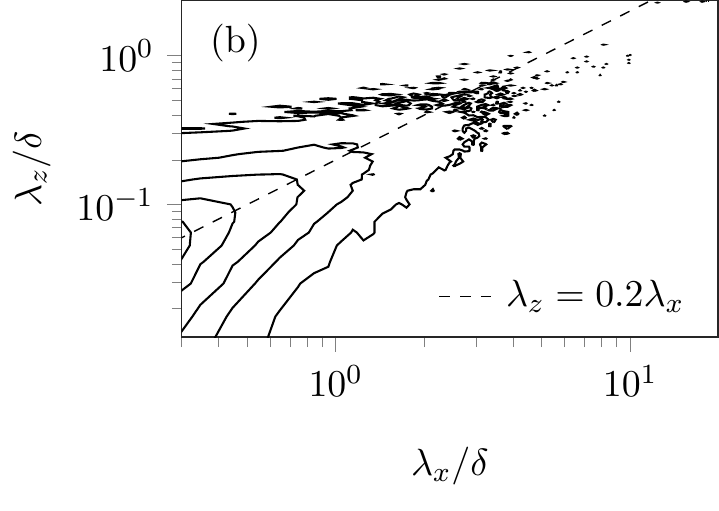}
    \end{minipage}

    \begin{minipage}{0.45\columnwidth}
        \includegraphics{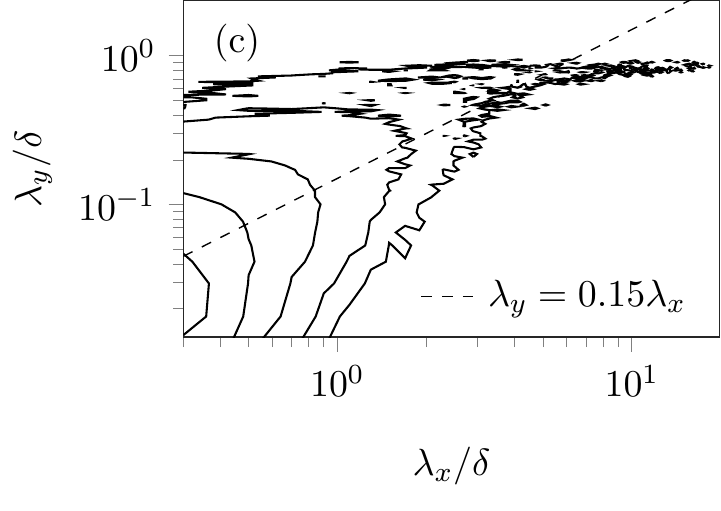}
    \end{minipage}
    \begin{minipage}{0.45\columnwidth}
        \includegraphics{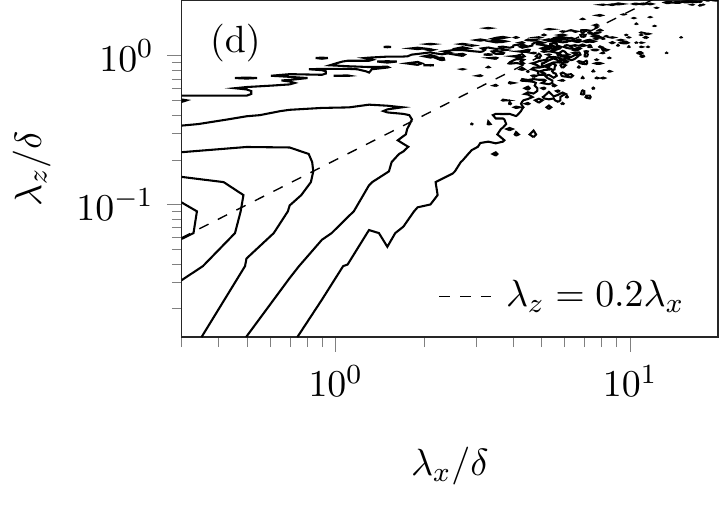}
    \end{minipage}

    \caption{Joint PDFs of streamwise wall-normal sizes $P(\lambda_x/\delta,\lambda_y/\delta)$ and streamwise spanwise size $P(\lambda_x/\delta,\lambda_z/\delta)$ of the detected structures.
    Areas inside the contour lines correspond to 99\%, 90\%, 75\%, 50\% and 25\% of the detected structures. An indicative ratio between the two sizes of the joint PDFs are given with dashed lines.
    Figures (a) and (b) are based on the binary volume $\mathbb{B}^{\ominus}$  while (c) and (d) are based on $\mathbb{B}^{\oplus}$} 
\label{fig:pdf_2D} 
\end{figure}

The aspect ratios of the attached structures obtained by 3D detection are investigated via joint PDFs of their length $P(\lambda_x/\delta, \lambda_y/\delta)$ and $P(\lambda_x/\delta, \lambda_z/\delta)$ (Figure~\ref{fig:pdf_2D}). The results show a clear trend for the shape of the detected structures with an average spanwise size around $20\%$ of the streamwise length for both low and high momentum ones. On the other hand, average height is slightly different for the low momentum and high momentum structures with $10\%$ and $15\%$ of the streamwise length respectively.  

Structures of the high-speed streamwise velocity fluctuations are limited with the top of the boundary layer while the height of the low-speed velocity structures can be taller than $\delta$. These results are compatible with the results given by Lozano-Dur{\'a}n \etal~\cite{lozanoduran2012} for attached quadrants as they showed that $Q2$ structures (related with the low-speed streamwise velocity fluctuations) are taller than $Q4$ structures.
Additionally, attached streamwise velocity fluctuation structures are spatially self-similar like attached quadrant structures detected by Lozano-Dur{\'a}n and co-workers with a quite similar streamwise/wall-normal aspect ratio.  
Very-large-scale structures appearing as the overhangs of the joint PDF $P(\lambda_x,\lambda_y)$ are not following the same self-similarity scaling. Lozano-Dur{\'a}n has also indicated this behavior for attached quadrant structures (See Figure 5 of ~\cite{lozanoduran2012})

\begin{figure}
        \begin{minipage}[b]{0.45\columnwidth}
            \includegraphics{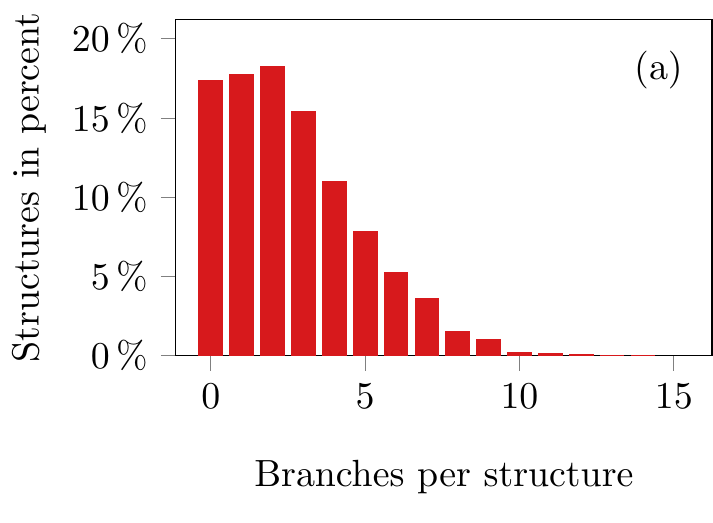}
        \end{minipage}
        \begin{minipage}[b]{0.45\columnwidth}
            \includegraphics{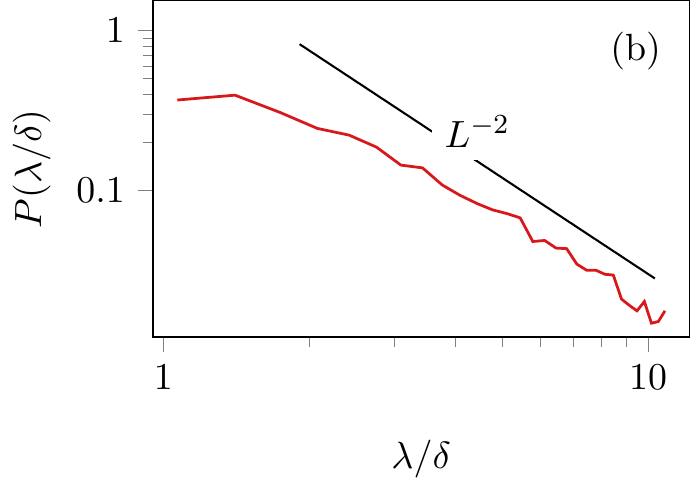}
        \end{minipage}

        \caption{Histogram of the number of branches having a length of at least $10 \%$ of its main curve for the structures longer than $4\delta$ (a) and PDF of the structure lengths computed from the main curve of the skeleton, $P(\lambda/\delta)$ (b). The binary structures are extracted by Eq. (\ref{eq:thr}) with  $C_{thr}=1.0$. $8500$ main curves are used for the PDF.}
        \label{fig:pdf_lx_skel} 
    \end{figure}

The spatial self-similarity of the PDF of streamwise velocity fluctuation structures of size up to 2 to 3 $\delta$ (Figure \ref{fig:pdf_2D}) should be interpreted with caution. Indeed, the detailed view provided by skeletons of the same structures do not strongly support the self-similar shape of these structures which appear to have complex multi-branch shapes rather than a simple core shape (Figure \ref{fig:skeleton_3d_single}).
The complex shape of the long structures can be characterized by counting the number of branches with a significant length. This measure, only accessible from the skeletons, expresses the intricate 3D shape of the structures. The histogram of the number of branches with a length of at least $10 \%$ of their main curve for the longest structures (longer than $4\delta$) is given Figure~\ref{fig:pdf_lx_skel} (a). Most of the retained large-scale structures have at least one significant branch and more than $40\%$ of them have even more than $3$ significant branches. Such multi-branch structures are difficult to characterize with a single streamwise length and a mean angle with respect to the horizontal plane.
The largest branches could be interesting to investigate as well. Unfortunately, this statistics would require a larger number of long structures to convergence statistics and are therefore out of reach with our DNS database.

Additional quantitative results of the structures can be extracted from their skeletons. The total length of a structure is computed differently, by extracting the main (longest) curve from its skeleton. Note that the direction of the skeleton is not necessarily aligned with the streamwise direction, unlike the size of the bounding box, $\lambda_x$. The PDF of the main curve lengths, $P(\lambda/\delta)$ computed from the skeletons is shown in Figure~\ref{fig:pdf_lx_skel}. Structures up to $\mathcal{O}(10\delta)$ are detected in agreement with some previous results, and the distribution exhibits the same $\lambda^{-2}$ slope which is equivalent to the statistics computed from the streamwise size of the boxes circumscribing the 3D binary objects. 

\begin{figure}

    \includegraphics{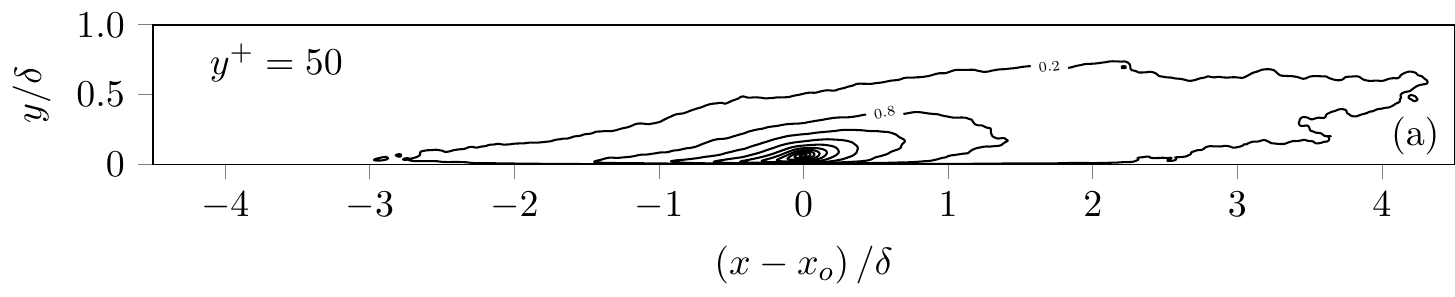}

    \includegraphics{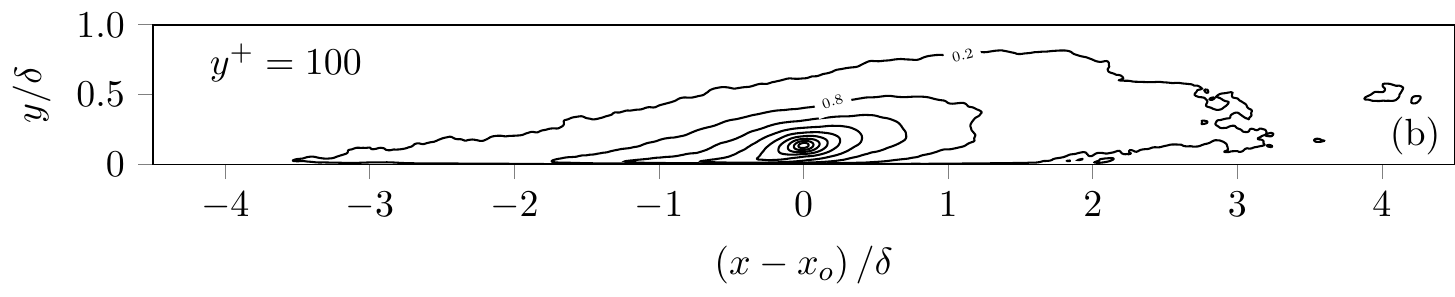}

    \includegraphics{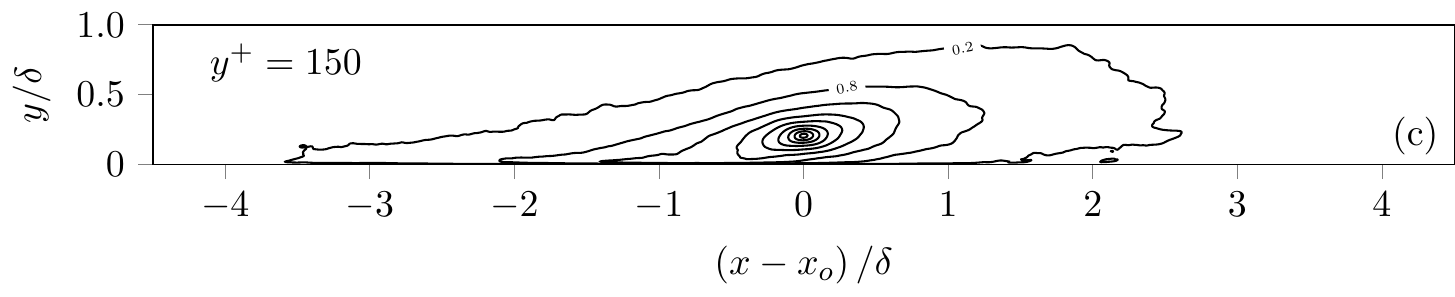}
    \caption{Two-point spatial correlation function of the streamwise velocity fluctuations $\langle u(x{-}x_o,y{-}y_o)u(x,y)\rangle$ conditioned by $\vert u(x_o,y_o) \vert > C_{thr} \sigma_u^{100^+} $ at wall distances (a) $y_o=50^+$, (b) $y_o=100^+$ and (c) $y_o=150^+$ where $x_o$ is the streamwise position such that $Re_\theta=2068$ and $C_{thr}=1$. The condition $\vert u(x_o,y_o) \vert  > C_{thr} \sigma_u^{100^+}$ represents only $32\%$ of volume but $80\%$ of the streamwise kinetic energy.}%
    \label{fig:2p_corr_inStr}
\end{figure}

The mean angle of the large-scale structures can be estimated from two-point correlations (Figure~\ref{fig:2p_corr}). However, since two-point correlations are computed with all possible fixed points, they do not only represent the most intense streamwise large-scale structures. Therefore, conditional two-point spatial correlations computed with the fix points inside the detected binary object  ($\mathbb{B}^{\ominus}$ and  $\mathbb{B}^{\oplus}$) are also provided (Figure~\ref{fig:2p_corr_inStr}). 
Using the same normalization $\langle u^\prime u^\prime \rangle (x_o, y_o)$ than the standard two-point correlations (Figure~\ref{fig:2p_corr}) longer iso-contours extending up to $7\delta$ are observed. 
These conditioned statistics are more representative of the average shape and size of the most energetic large-scale streamwise structures than the standard correlations. 

\begin{figure}
    \begin{minipage}[t]{0.495\columnwidth}
        \includegraphics{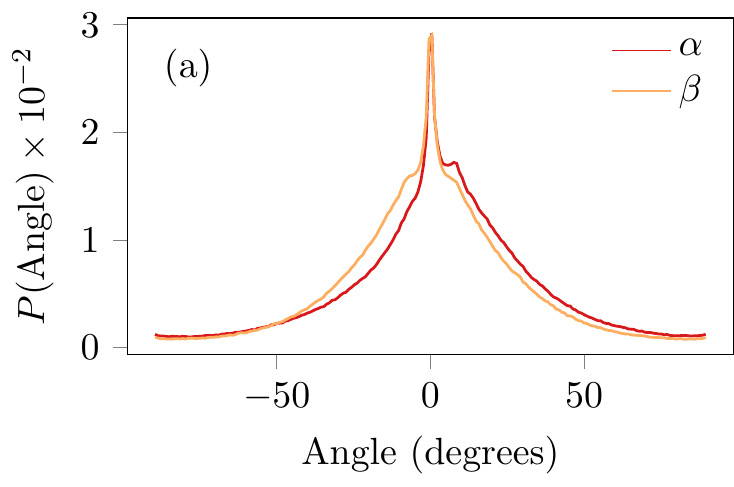}
    \end{minipage}
    \begin{minipage}[t]{0.495\columnwidth}
        \includegraphics{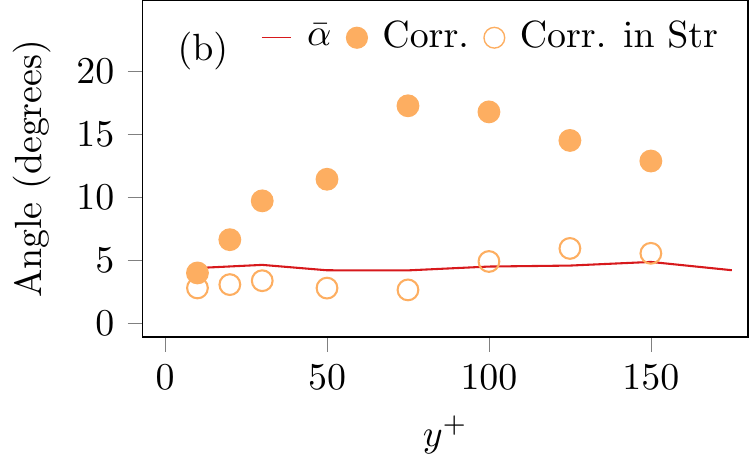}
    \end{minipage}
    \caption{\label{fig:mean_ang_skel} (a) PDFs of the pitch ($\alpha$) and yaw ($\beta$) angles from the main curve of the skeletons. (b) Mean pitch angle ($\alpha$) as a function of wall distance extracted from skeleton in comparison with the mean angles detected from the two-point spatial correlations functions (Figures~\ref{fig:2p_corr} and \ref{fig:2p_corr_inStr}). The mean angle from two-point correlation function considered as the slope of the longest line between the fixed point and the $80 \%$ isocontour.}
\end{figure}

An advantage of the skeletons is that the angles at any point of the curves can efficiently be computed for each structure as the curves are defined with sub-pixel accuracy. 
The skeleton angles are extracted only on the main curves (the black curve in Figure~\ref{fig:skeleton_3d_single}) to exclude the statistics of the branches which may have different statistics and a different meaning.
The angles of the large structures (with the main curve longer than $\delta$) are defined between the unit vector $\bar{\textbf{x}}$ and the projections of the displacement vector of two consecutive points of a curve on the XZ-plane (pitch angle, $\alpha$) and the XY-plane  (yaw angle, $\beta$).
Their distributions are given in Figure~\ref{fig:mean_ang_skel} (a). 
The large tails of the PDF indicate that the main curves move up and down with a large distribution of angles as noticed in the Figure~\ref{fig:skeleton_3d_single} for a single structure.
 The distribution of the pitch angle ($\alpha$) is right-skewed, in agreement with the positive average angle seen in two-point spatial correlations. Both pitch and yaw angles are widely distributed with no clear preferable values. The peak around very small angles and the short plateau around the small positive value of the distributions are the consequence of irregular oscillations of the skeletons.

The mean pitch angle ($\alpha$) of the structures is evaluated from their main curves as a function of the wall distance. It is presented in comparison with the angle extracted from the isocontours of the standard (Figure~\ref{fig:2p_corr}) and conditioned (Figure~\ref{fig:2p_corr_inStr}) two-point correlations. The statistics from skeletons show that the detected large-scale structures of streamwise velocity fluctuations move away from the wall with an average pitch angle of $5^\circ$, almost independently from the wall-normal distances (Figure~\ref{fig:mean_ang_skel}). The conditional two-point correlations using only extreme streamwise fluctuations lead to a similar estimation of the mean pitch angle ($\alpha$) while standard two-point spatial correlations lead to a higher average value as it also includes the information from regions with weak fluctuations but also from strong detached structures. 

\begin{figure}
  \centering
  \begin{minipage}[b]{0.495\columnwidth}
      \includegraphics{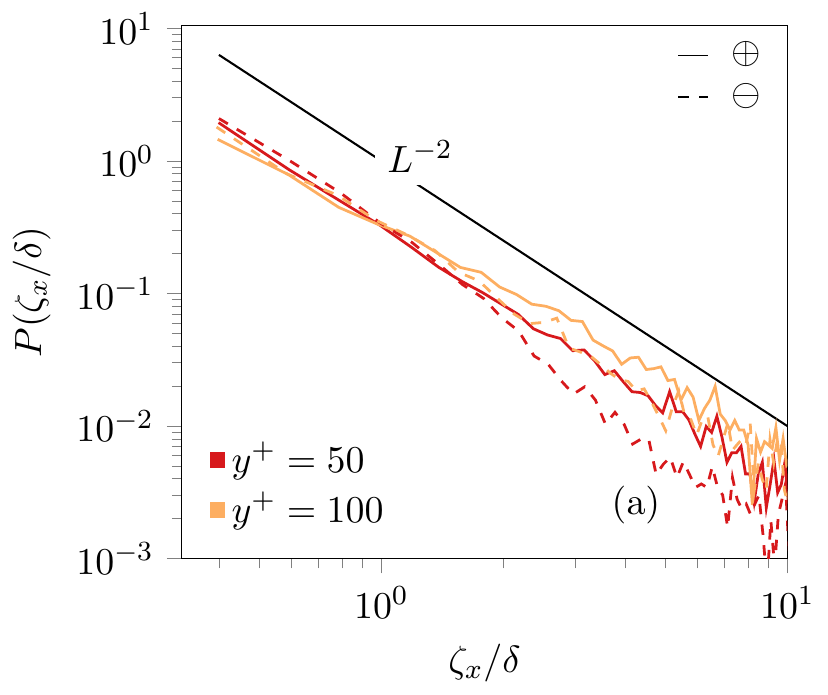}
  \end{minipage}
  \begin{minipage}[b]{0.495\columnwidth}
      \includegraphics{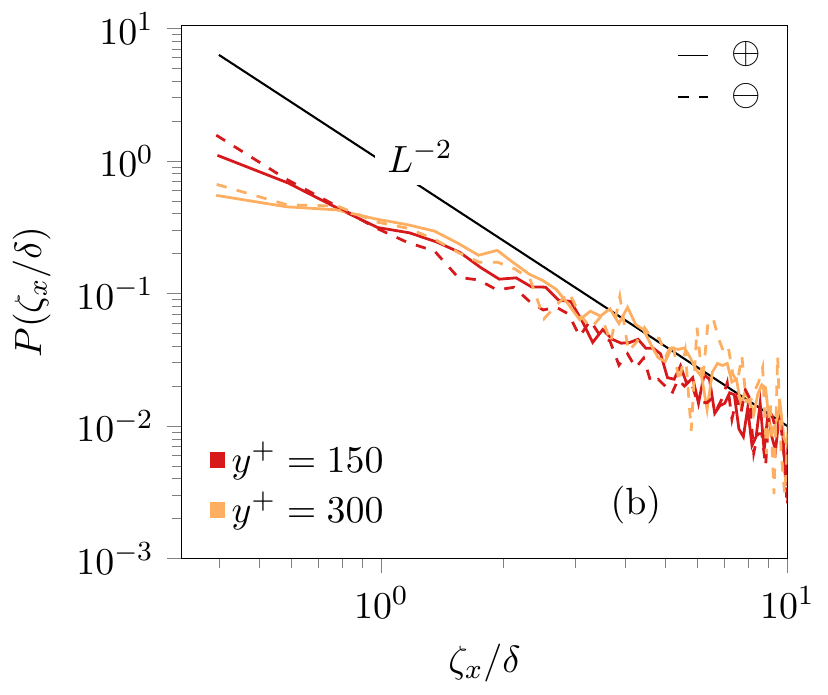}
  \end{minipage}

    \caption{\label{fig:pdf_zeta_x} PDF of the structure lengths at particular wall-distances using 3D detection from negative velocity fluctuations (dashed lines) and positive velocity fluctuations (continuous lines).}
\end{figure}

As discussed in the introduction, the LSMs near the wall have been interpreted as being responsible for the $k_x^{-q}$ scaling range with $q \simeq 1$ of the turbulence spectrum. Following the analysis of Srinath \etal~\cite{srinath2017} on TBL experimental data, the turbulence intensity of the detected streamwise large-scale structures at a certain wall distance is evaluated. The length of the structure in the streamwise direction at a particular wall distance $\zeta_x$ is not necessarily equal to $\lambda_x$ because structures have irregular shapes and may include holes due to a small region of streamwise fluctuating velocity lower than the selected threshold. Similar to the distribution of $\lambda_x$ (Figure~\ref{fig:pdf_lambda_x}), $\zeta_x$ also demonstrates a fairly good $-2$ power law up to $y^+=150$ ($\simeq 0.2\delta$) as shown in Figure~\ref{fig:pdf_zeta_x}. 

\begin{figure}
    \begin{minipage}[t]{0.3\columnwidth}
        \includegraphics{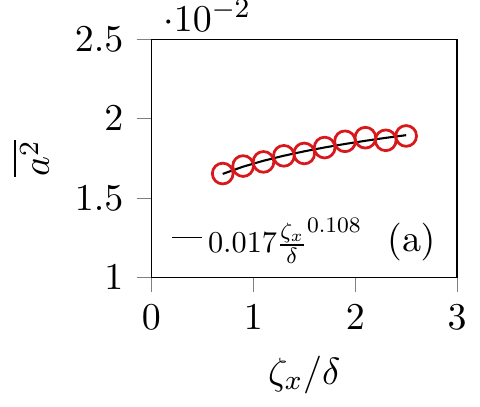}
    \end{minipage}
    \begin{minipage}[t]{0.3\columnwidth}
        \includegraphics{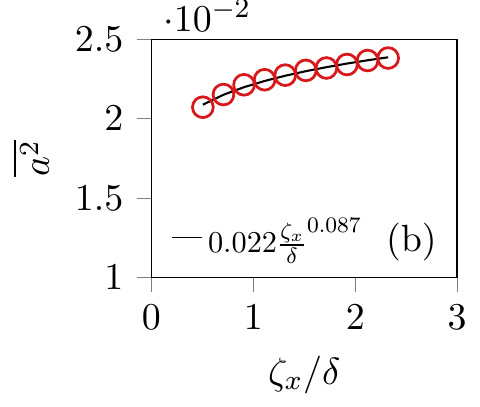}
    \end{minipage}
    \begin{minipage}[t]{0.33\columnwidth}
        \includegraphics{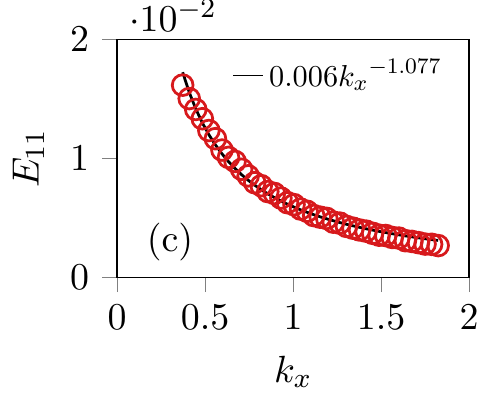}
    \end{minipage}

   \begin{minipage}[t]{0.3\columnwidth}
        \includegraphics{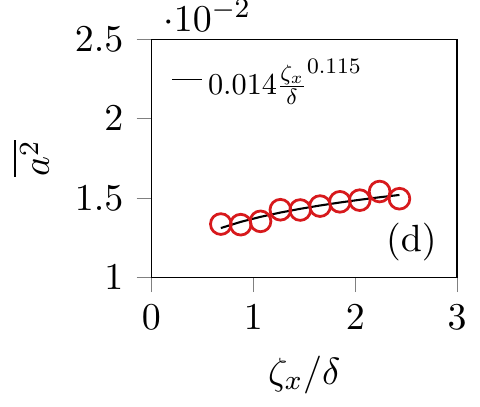}
    \end{minipage}
    \begin{minipage}[t]{0.3\columnwidth}
        \includegraphics{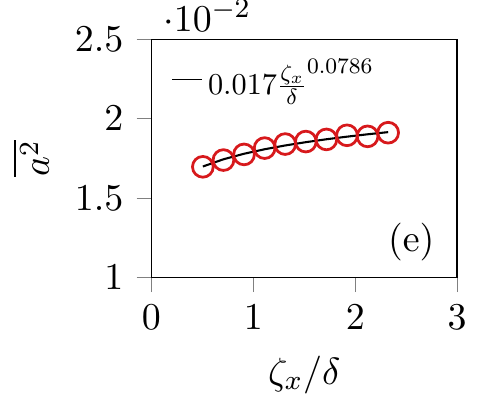}
    \end{minipage}
    \begin{minipage}[t]{0.33\columnwidth}
        \includegraphics{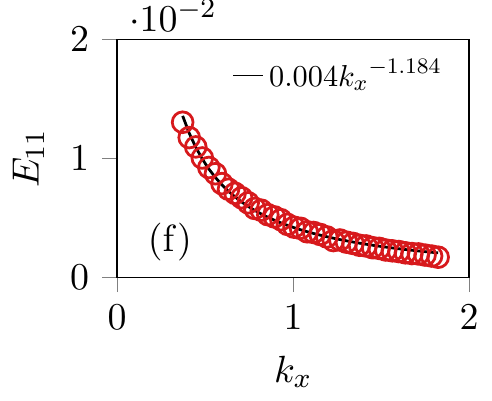}
    \end{minipage}

    \begin{minipage}[t]{0.3\columnwidth}
        \includegraphics{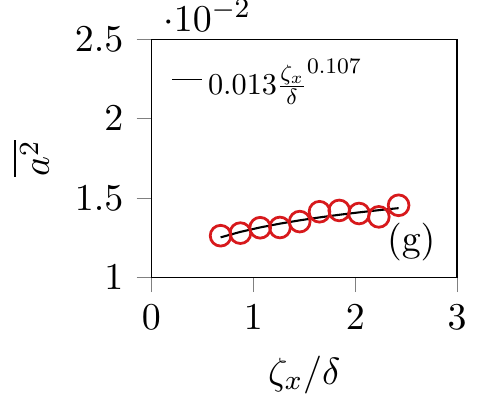}
    \end{minipage}
    \begin{minipage}[t]{0.3\columnwidth}
        \includegraphics{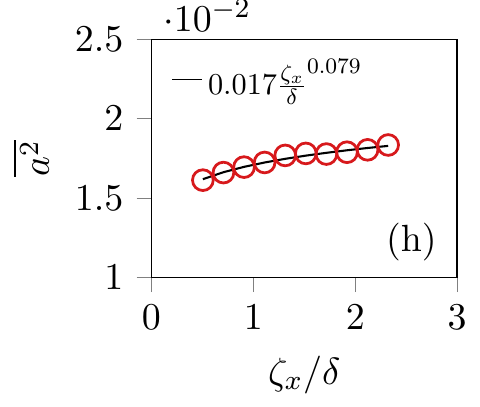}
    \end{minipage}
    \begin{minipage}[t]{0.33\columnwidth}
        \includegraphics{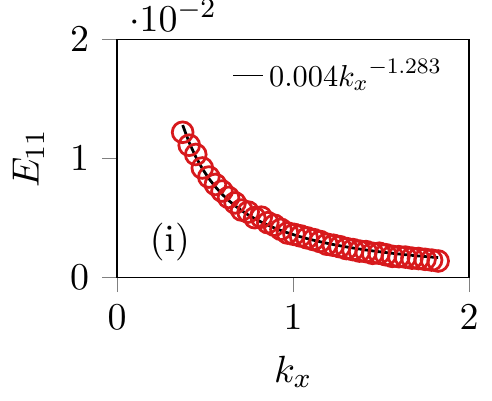}
    \end{minipage}
    \caption{\label{fig:tp_model_fits} Average mean squared streamwise velocity, $\overline{a^2}$ as function of structure length in 3D structures ((a), (d), (g)) and in 2D structures ((b), (e), (h)). Streamwise energy spectra ((c), (f), (i)) at three wall distances: $y^+ = 50,\,100,\,150$ (from top to bottom). All statistics are computed for $Re_\theta = 2068$.}
\end{figure}

The fluctuating streamwise velocity averaged inside the detected structures is defined as $\overline{a^2}$. 
It has been computed per line of the detected structures in 2D as in the Srinath \etal~\cite{srinath2017} while the results from 3D detected structures also averaged in the spanwise direction. 
Therefore, the distributions of $\overline{a^2}$ at a particular distance from the wall as a function of $\zeta_x$ (Figure~\ref{fig:tp_model_fits}) are not exactly the same for the 2D, and 3D structures as the structures are not identical.

A $\zeta_x^{p}$ power law fit of $\overline{a^2}$ is evaluated on the same limited range of structure length as in Srinath \etal~\cite{srinath2017} (from $0.5\delta$ to $2.5\delta$) where the energy spectra exhibits a decent $k_x^{q}$ region (with $q \simeq 1$) and where the distribution of structure lengths follows a ${-}2$ power law (Figure~\ref{fig:pdf_zeta_x}). 
Srinath \etal~\cite{srinath2017} have shown that, for the experimental data at higher Reynolds number, power law exponents $p$ and $q$ are such that  $p + q$  is very close to $-1$. 
They demonstrated such results for a range of distances from the wall ($y=50^+$) up to $0.1 \delta$ with a special value of $p=0$ near $y^+=150$ leading to an almost perfect $k_x^{-1}$ slope for the streamwise energy spectra at this particular wall distance. 
In the present study, the $q$ exponent of the streamwise energy spectra is decreasing when moving from the wall but the slope of $\overline{a^2}$ stays fairly constant as compared to the results of \citet{srinath2017}. 
This can be related to the moderate Reynolds number of the present study as $y^+=150$ corresponds to $0.2\delta$ and the relationship $p+q=1$ has been observed up to $0.1\delta$ on the experimental data. 
However, the power law fits of both distributions of $\overline{a^2}$ on 2D and 3D detected structures have a small positive exponent $p \simeq 0.1$ at $y^+=50$ such as $p + q$ is close to $-1$. 
For example, $p=0.115$ for 3D labeled structures and $q=-1.184$ at $y=100^+$ leading to $p+q=-1.069$.
Despite the uncertainties of the fit due to the limited $k_x^{q}$ range of $E_{11}$ at lower Reynolds number, the present results support the findings of Srinath \etal \cite{srinath2017} both with the 2D and 3D structures.

\section{\label{sec:conc}Conclusion}

A large database of 3D fields and 2D time-resolved data were collected from a new DNS for TBL flow at moderate Reynolds number. Spectral analysis shows that the streamwise energy spectrum is compatible with a $k_x^{-q}$ scaling at large scales with $q\simeq 1$ near $y^+=100$ with increasing value of $q$ when moving from the wall. Even though, the second peak (or plateau) of ${u^\prime}^+$ is not clear at this Reynolds number, a reference value of $\sigma_u^{100^+}$ at $y^+=100$  of the streamwise velocity fluctuation is used in the detection procedure of the large-scale structures. 

Structures are detected with a simple thresholding algorithm (Eq.~\ref{eq:thr}). 
The attached large-scale structures represent a substantial part of the streamwise velocity fluctuations. 
Detached structures with smaller fluctuations can probably fulfill the gap in the outer region. 

The 3D shape of the structures is also studied. Two different extractions of the structures were compared. The analyses of the 2D structures are comparable to the results given by Srinath \etal~\cite{srinath2017} on PIV data at higher Reynolds number. Moreover, a second method which benefits from having 3D numerical data is also applied, and the results are compared with the first method to make the connection with previous findings from experimental data. It is shown that 2D and 3D extractions may lead to different structure sizes due to the 3D nature of the structures. Information in the spanwise direction is meaningful as the longest structures also have a more complex topology. Representing the complex 3D shape of this structures with a single length scale in each direction is not obvious.
A specific skeletonization algorithm is used to simplify the 3D binary image of a single detected structure.
Skeletons consist of a set of well-resolved lines which can be used to better characterize them regarding length, complexity, and angles. This simplification algorithm does not rely on lines of extrema which may be not representative of the full structure but rather defines curves based on its shape.

Distributions of the streamwise lengths of detected structures are investigated based on the results of both bounding boxes and skeletons. The results show that the distribution of streamwise lengths of the large-scale structures follows the same power law with a slope close to the $-2$ in agreement with the findings of  Srinath \etal~\cite{srinath2017}. It seems that the streamwise length distribution of the attached streamwise velocity fluctuation is quite robust as the same $-2$ scaling is found with different detection methods (2D and 3D). Joint PDFs $P(\lambda_x/\delta, \lambda_y/\delta)$ and $P(\lambda_x/\delta, \lambda_z/\delta)$ are given to provide 3D statistical descriptions of the detected structures.
The results reveal some self-similarity between structures of different size but on a limited range due to the moderate Reynolds number of the DNS data. 
A noticeable agreement between the shapes of the attached streamwise velocity structures detected in this study and attached quadrant structures detected by Lozano-Dur{\'a}n \etal~\cite{lozanoduran2012} is also observed . 

The preferred angle of the detected structures is also investigated. 
Based on the main (longest) curve of the skeletons, a constant mean value of the angle is observed for all wall distances which is different from the results extracted from the standard two-point spatial correlations. 
Even though, measurements on the main curve of the skeletons suggest that large-scale wall attached structures follow an upward trend around $5^\circ$ on average along the full boundary layer thickness, a model of an elongated structure with a single positive angle with respect to the horizontal plane is not representative of the real shape of individual structures. 

Thresholding of the streamwise velocity fluctuations with a single parameter reveals the complex shapes of these streamwise structures and the skeletonization procedure is used to quantify their features. Their complexity was asserted by counting the number of significant branches of a structure. It is shown that most of the large structures ($> 4\delta$) have one or more significant branches. 

One can conclude that the characterization of the large-scale motions with a single length scale and an average angle may be challenging. However, the present analyses show that the scaling of the attached streamwise structures length and turbulence intensity are not too sensitive to the selected detection methods.

\section*{Acknowledgments}

This work was granted access to the HPC resources of IDRIS and CINES under the allocation 021741 made by GENCI (Grand Equipement National de Calcul Intensif). The authors are grateful to S. Laizet for providing the code Incompact3d used for the numerical simulation and to Prof. W.K. George for useful discussions.

\appendix
\section{Skeletonization\label{app:skel}}

\begin{figure}
		\begin{centering}
	\includegraphics{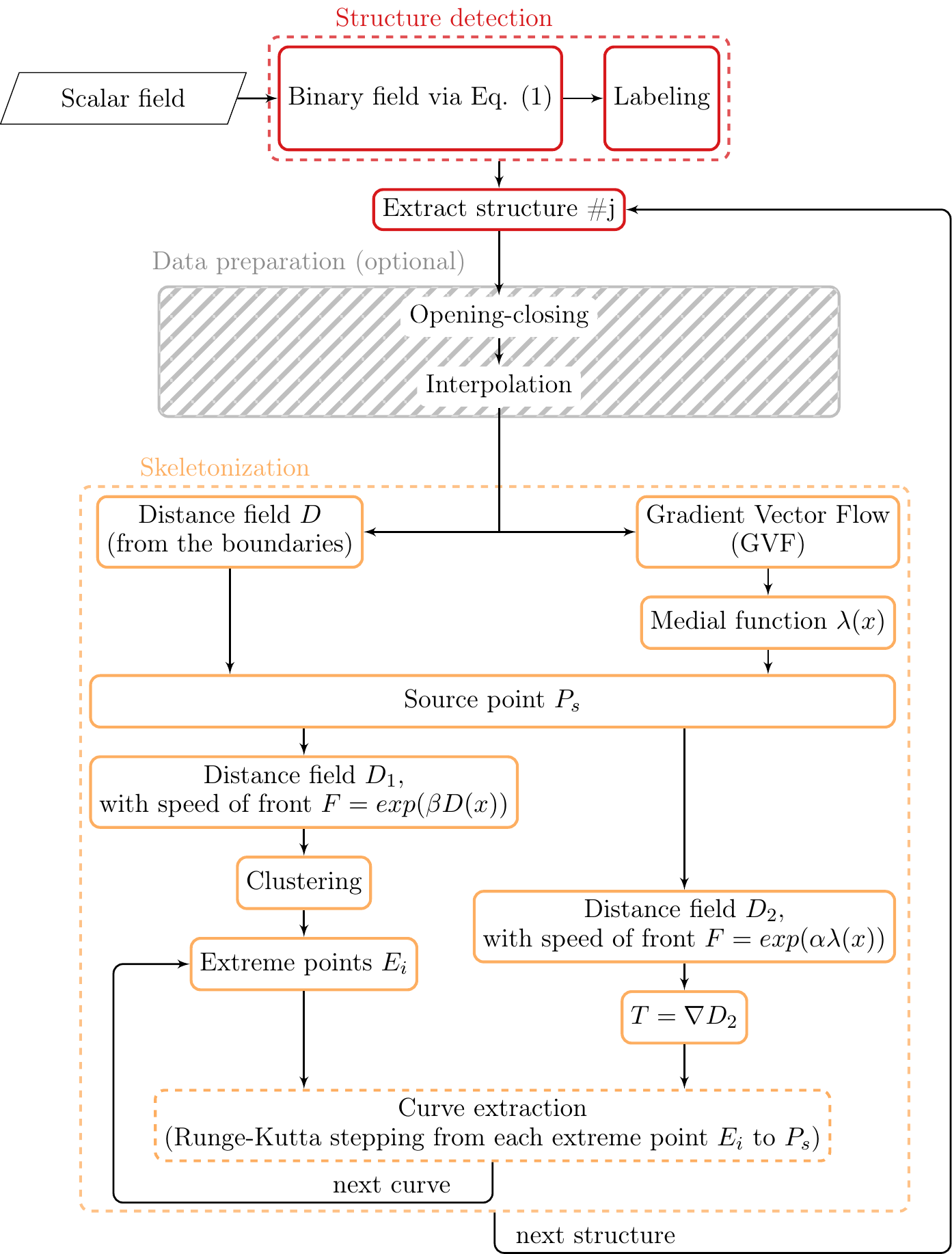}
			\end{centering}
        \caption{Graphical presentation of the work-flow to extract the skeleton's curves for a single 3D spatially resolved data. The red steps correspond to the structure detection method used priorly to the bounding box statistics.}
		\label{fig:code_inf_mat}
\end{figure}

The curve extraction framework proposed by Hassouna \etal~\cite{hassouna2009}
along with structure detection steps is abstracted in Figure~\ref{fig:code_inf_mat}.

\noindent

The main steps of the method can be summarized as follows:

\begin{enumerate}
    \item Compute the first distance field $D$ from boundaries solving an Eikonal equation,
    \item Compute the gradient vector flow (GVF) based medial function $\lambda(x)$ that points out toward the source point of the object,
    \item Detect the source point $P_s$ defined as $P_s=argmax(\lambda D)$
    \item Propagate a $\beta$-front from $P_s$ and solve a new Eikonal equation to get a new distance field $D_1$;
    \item Discretize $D_1$ (clusterize) and construct level-set-graph, which represent the connection between adjacent clusters,
    \item Detect extreme and merge points for appropriate clusters,
    \item Propagate a $\alpha$-front from $P_s$ and solve a new Eikonal equation to get a new distance field $D_2$,
    \item Extract curves by backtracking from extreme points following the negative gradient of $D_2$ (stops at $P_s$ or on a previously extracted part of the skeleton).
\end{enumerate}

Two optional procedures (opening-closing and interpolation) can be applied after the extraction of the 3D structure if the structure exhibit some regions which are not sufficiently resolved (a minimum resolution of $5 \times 5 \times 5$ is necessary for the healthy progress of Runge-Kutta stepping to extract a curve). 

\clearpage
\bibliography{bibtex_file}

\end{document}